\documentclass[pdflatex,sn-mathphys-ay]{sn-jnl}

\usepackage{graphicx}%
\usepackage{multirow}%
\usepackage[normalem]{ulem}
\usepackage{amsmath,amssymb,amsfonts}%
\usepackage{amsthm}%
\usepackage{mathrsfs}%
\usepackage[dvipsnames]{xcolor}
\usepackage{textcomp}%
\usepackage{manyfoot}%
\usepackage{booktabs}%
\usepackage[ruled, vlined, longend]{algorithm2e}%
\usepackage{algpseudocode}%
\usepackage{listings}%
\usepackage{lmodern}
\usepackage{helvet}
\usepackage{multirow}
\usepackage{colortbl}
\usepackage{tabularx}
\usepackage{caption}
\usepackage{subcaption}
\usepackage{chngcntr}
\counterwithout{figure}{section}
\graphicspath{{../pdf/}{../jpeg/}{../svg/..}{../eps_tex/..}}
\newcommand{\code}{\texttt}
\usepackage{mathtools}
\usepackage[inkscapelatex=false]{svg}
\usepackage{tikz}
\usetikzlibrary{matrix, positioning}
\usetikzlibrary{tikzmark}

\newcommand{\rev}[1]{\textcolor{blue}{\uwave{#1}}} 
\renewcommand{\rev}[1]{#1}

\usepackage{titlesec}
\titleformat{\paragraph}
{\normalfont\normalsize\bfseries}{\theparagraph}{1em}{}
\titlespacing*{\paragraph}
{0pt}{3.25ex plus 1ex minus .2ex}{1.5ex plus .2ex}
\let\oldhat\hat 
\renewcommand{\vec}[1]{\mathbf{#1}} 
\renewcommand{\hat}[1]{\oldhat{\mathbf{#1}}}

\begin{document}

\title[HQSL]{A Hybrid Quantum Neural Network for Split Learning}

\author*[1,2]{\fnm{Hevish} \sur{Cowlessur}}\email{hcowlessur@student.unimelb.edu.au}

\author[2]{\fnm{Chandra} \sur{Thapa}}\email{chandra.thapa@data61.csiro.au}

\author[1]{\fnm{Tansu} \sur{Alpcan}}\email{tansu.alpcan@unimelb.edu.au}

\author[2]{\fnm{Seyit} \sur{Camtepe}}\email{seyit.camtepe@data61.csiro.au}
\affil[1]{\orgdiv{Department of Electrical and Electronic Engineering}, \orgname{University of Melbourne}, \orgaddress{\city{Parkville}, \postcode{3010}, \state{VIC}, \country{Australia}}}

\affil[2]{\orgdiv{Data61}, \orgname{CSIRO}, \orgaddress{\city{Sydney}, \state{NSW}, \country{Australia}}}

\abstract{

Quantum Machine Learning (QML) is an emerging field of research with potential applications to distributed collaborative learning, such as Split Learning (SL). SL allows resource-constrained clients to collaboratively train ML models with a server, reduce their computational overhead, and enable data privacy by avoiding raw data sharing.
Although QML with SL has been studied, the problem remains open in resource-constrained environments where clients lack quantum computing capabilities.
Additionally, data privacy leakage between client and server in SL poses risks of reconstruction attacks on the server side. To address these issues, we propose Hybrid Quantum Split Learning (HQSL), an application of Hybrid QML in SL. HQSL enables classical clients to train models with a hybrid quantum server and curtails reconstruction attacks. Additionally, we introduce a novel qubit-efficient data-loading technique for designing a quantum layer in HQSL, minimizing both the number of qubits and circuit depth. Evaluations on real hardware demonstrate HQSL's practicality under realistic quantum noise.
%
%
Experiments on five datasets demonstrate HQSL’s feasibility and ability to enhance classification performance compared to its classical models. 
Notably, HQSL achieves mean improvements of over 3\% in both accuracy and F1-score for the Fashion-MNIST dataset, and over 1.5\% in both metrics for the Speech Commands dataset. We expand these studies to include up to 100 clients, confirming HQSL's scalability.
Moreover, we introduce a noise-based defense mechanism to tackle reconstruction attacks on the server side.
%
Overall, HQSL enables classical clients to train collaboratively with a hybrid quantum server, improving model performance and resistance against reconstruction attacks.
}
\maketitle
\section{Introduction}\label{sec1}
Quantum Machine Learning (QML) is a rapidly evolving research area at the intersection of quantum computing and Machine Learning (ML). A few examples of quantum analogs to some classical ML algorithms are Quantum Support Vector Machines \citep{PhysRevLett.114.140504}, Quantum Neural Networks (QNN) \citep{jia2019quantum} and Quantum Principal Component Analysis \citep{lloyd2014quantum}. These leverage the unique properties of qubit systems, such as superposition and entanglement, to study the potential of improving ML tasks using quantum computing. 
While fault-tolerant quantum computers are still a few years away, a practically feasible route for addressing real-world problems in the Noisy-Intermediate Scale Quantum (NISQ) era \citep{Preskill2018quantumcomputingin} is by adopting a hybrid approach. 
The hybrid approach integrates classical deep neural networks with the quantum layers, i.e., Hybrid Quantum Neural Network (HQNN) \citep{liang2021hybrid, liu2021hybrid}. However, there are few works on HQNN, and it remains open in various aspects, including a resource-constrained distributed learning environment such as ML in the Internet of Things domain. 

In distributed collaborative learning, Split Learning (SL)~\citep{vepakomma_split_2018} is a popular technique in resource-constrained environments where clients have low computing power. The ML model is split into client and server components, where the client trains the initial few layers on local data, and then the server continues training the remaining layers. SL allows collaborative training without sharing raw data. In the quantum computing domain, Quantum Split Learning (QSL) splits a QNN to leverage SL advantages~\citep{yun_quantum}. However, in resource-constrained NISQ era environments, it is crucial to allow clients (no quantum computing capabilities) to compute in the classical domain. This motivates HQNN in SL which has not been studied before. As SL can suffer from data privacy leakage\footnote{Data privacy leakage in split learning is the risk of private information being revealed or reconstructed from exchanged information during collaborative training, despite not sharing raw data directly.}, there is a risk of reconstruction attacks on the server side if the server is honest but curious~\citep{joshi_performance_2022, li_label_2022}. Moreover, this security aspect has not been investigated in HQNN with SL. 

We present a novel (to the best of our knowledge, the first) application of HQNN to SL, which we call Hybrid Quantum Split Learning (HQSL), and investigate its resistance against reconstruction attacks due to data privacy leakage.
In a simple HQSL setup, the HQNN model is divided into two parts: the client-side model, which has only classical neural network layers, and the server-side model, which consists of a quantum layer followed by classical neural network layers. This design reflects realistic deployment settings, where clients are typically resource-constrained (e.g., edge or IoT devices) and can only support a small number of lightweight classical layers, while compute-intensive operations and all quantum processing are offloaded to the more capable server.
The quantum layer consists of a quantum circuit that converts classical inputs into quantum states, performs quantum computations, and outputs classical data. Considering the limitations of NISQ computers, which restrict the number of qubits and circuit size, we employ a low-depth 2-qubit quantum circuit within the quantum layer. This also ensures our simulations have reasonable runtimes. 
To allow a relatively larger dimension of classical data inputs into our quantum circuit, we propose a qubit-efficient data-loading technique. To tackle data privacy leakage in HQSL, we integrate a noise layer based on Laplacian distribution into the client-side model. In contrast to existing works on the Laplacian noise layer, we tune the noise parameters considering the rotational properties of the quantum layer's encoding gates. This paper is also the first, to the best of our knowledge, to study reconstruction attacks in the hybrid quantum-classical domain.

\noindent
Overall, our contributions are the following: 
\begin{enumerate}
   \item \textit{Feasibility and performance of HQSL with a single client:} 
    We propose a novel HQSL architecture by combining SL with HQNN. This allows resource-limited classical clients to collaboratively train HQNN models with a hybrid quantum server featuring a 2-qubit quantum circuit layer followed by classical dense layers.
    Despite the small size of the quantum circuit, we successfully showcase a modest improvement in the test accuracy and F1-score over the classical counterpart of HQSL by conducting extensive experiments with five standard datasets as a proof of concept. This demonstrates the feasibility and superior performance of HQSL over SL. We further validate the practicality of HQSL deployment by evaluating its performance on multiple real quantum devices and noisy simulators.

    \item \textit{Quantum Circuit Design:} We devise a qubit-efficient data-loading scheme, consisting of only 2 qubits and 3 layers of encoding and parameterized gates corresponding to a 3-dimensional input. Through experimental analyses on the Fashion-MNIST dataset \citep{xiao2017fashionmnist}, we show that HQSL with our proposed circuit in its quantum layer features a smaller simulation runtime and better classification performance than deeper quantum circuits.
    \item \textit{HQSL with multiple clients:} 
    We extend our studies of HQSL by including up to 100 clients, illustrating its scalability. This advancement highlights HQSL's ability to support numerous classical clients in collaborative HQNN model training.
    \item \textit{Resistance to reconstruction attacks: }We study the resistance of HQSL against reconstruction attacks happening at inference time on the server side and propose a quantum encoding gate-based Laplacian noise layer defense mechanism as a countermeasure. 
    A comparative study on reconstruction attacks in hybrid and classical settings demonstrates the higher resilience of HQSL compared to SL.
\end{enumerate}

The rest of this paper is organized as shown in Table \ref{tab:paper_org}.

\begin{table*}[htpb!]
    \centering
     \caption{Table of the organization of the paper in sections}
     \resizebox{0.95\columnwidth}{!}{%
    \begin{tabular}{|c|p{12cm}|} \hline 
         Section \ref{sec:Literature Review}& Related work\\ \hline 
         Section \ref{sec:Problem_statement}& Problem statement and research questions\\ \hline 
         Section \ref{sec: HQSL}& HQSL architecture, quantum circuit design and selection process, HQSL model training with single and multiple clients\\ \hline 
         Section \ref{sec:hqsl_exp}& Experiments and findings with HQSL involving single and multiple clients \\ \hline 
         Section \ref{sec:reconstruction}& Enhancing the resistance of HQSL against reconstruction attacks, proposed defense mechanism, experiments to test defense mechanism\\ \hline 
         Section \ref{sec: Conclusion}& Conclusion and future work\\ \hline
    \end{tabular}
    }
    \label{tab:paper_org}
\end{table*}

\section{\label{sec:Literature Review}Related Work}
This section reviews relevant works on Hybrid QML, SL, and Quantum Split Learning. 
We discuss the need for combining classical and quantum computing in the NISQ era and describe the SL scheme and its study in the pure quantum domain. 

\noindent
\textbf{\label{lit:hq_adv}Combining Classical and Quantum Resources.}
Currently available quantum computers exhibit noise, due to numerous factors, e.g., qubit decoherence and gate errors, constraining the complexity of quantum algorithms.
Thus, a common practice to leverage the power of quantum computations in the NISQ era is by combining it with classical methods. 
Hybrid QML models, e.g., Hybrid Quantum Neural Network (HQNN), are becoming a common area of research, where a quantum computer computes only part of the model. For instance, in HQNNs, classical and quantum computing nodes are concatenated, with the classical layers reducing high dimensional data to low dimensions to suit small quantum circuits that make up the quantum layer \citep{schetakis2022review, alam2021quantum}. This approach is particularly important for current-generation quantum hardware, which is constrained by the number of available qubits, circuit size and depth.
Moreover, in the Generative Adversarial Network (GAN) domain, notable works have proposed the use of quantum circuits with classical neural network layers that aid in pre-processing \citep{liu_hybrid_2021} and post-processing \citep{tsang_hybrid_2023} giving rise to hybrid quantum-classical GANs, which further highlight the versatility of quantum resources in hybrid architectures. In this work, we propose, for the first time to the best of our knowledge, yet another combination of quantum and classical resources in the form of \textit{an HQNN for the split learning domain}.

\noindent
\textbf{\label{lit:sl}Split Learning (SL) and the Data Privacy Leakage in SL.}
In the distributed learning domain, SL has addressed computational requirement problems with Federated Learning (FL). In FL, each client has to run the entire ML model, which can be too computationally intensive for resource-constrained clients \citep{thapa_splitfed_2022}. By offloading the compute-intensive parts of the model, SL enables more efficient resource allocation on the client devices, benefiting resource-constrained clients. 

However, while SL prevents the sharing of clients' raw data with the server, information leakage still occurs from smashed data\footnote{In split learning, smashed data refers to the intermediate output or activations produced by the last layer of the client-side portion of a split neural network.}. This is a concerning security issue because the smashed data contains latent information about the raw input data and can be used to stage an input data reconstruction attack \citep{abuadbba_can_2020} at the server side. In the literature, several countermeasures have been proposed against data privacy leakage in SL. Joshi et al. \citep{joshi_performance_2022} proposed increasing the number of layers in the client-side model. This reduces the mutual information, hence reducing the similarity between the smashed data and raw data. The downside of this technique is that it increases the computational overhead on the client side, which can be problematic for lightweight devices.
NoPeek, an information-theoretic measure to reduce information leakage, is another method suggested \citep{vepakomma_nopeek_2020} to reduce data privacy leakage. They do so by augmenting the classification loss function with the distance correlation function to reduce information leakage. However, a privacy-utility tradeoff is a possible consequence of NoPeek.
Other methods involve differential privacy \citep{gawron_feature_2022} and homomorphic encryption, which come at the cost of model performance degradation  \citep{shokri_privacy-preserving_2015} and too computationally intensive, respectively. In this work, we investigate the problem of data privacy leakage in the hybrid quantum split learning domain and \textit{propose a novel noise-based defense mechanism designed considering the rotational properties of single-qubit encoding gates.} The methodology, experiments, and results are presented in Section \ref{sec:reconstruction}.

\noindent
\textbf{\label{lit:qsl}Quantum Split Learning (QSL).}
In a pure QML framework, splitting a QNN has been trialed to examine the potential of QSL in preserving privacy and enhancing accuracy compared to Quantum Federated Learning (QFL) \citep{yun_quantum}. However, this work assumed a fault-tolerant regime whereby client devices are equipped with quantum capabilities--an unrealistic assumption given the current generation of quantum computers and resource-constrained clients. To address this, we propose a solution that allows clients to compute in the classical domain while the server operates in the hybrid quantum domain. This relaxation of the assumption ensures a more practical approach.
\section{\label{sec:Problem_statement}Problem Statement}
In this section, we define our research problem by breaking it down into specific research questions. This method enables us to systematically tackle the complexities of our study and offers a structured framework for our investigation.

\noindent
\textit{RQ 1: How can we leverage SL concepts to bring advantages of quantum computing to resource-constrained clients without quantum computing capabilities? -- addressed in Sections \ref{sec: HQSL}, \ref{sec:hqsl_exp}}.

Applying SL concepts in a pure QML model, e.g., splitting a QNN between a client and a server, assumes quantum computing capabilities on the client side. However, in resource-constrained environments, clients typically are resource-limited, e.g., IoT devices. Relaxing this assumption entails a client in the classical domain and a server equipped with quantum resources. The immediate sub-questions that arise are: 
(1) How can we model and train such a hybrid quantum split learning (HQSL) architecture? (2) How do we design a quantum circuit of low enough complexity for the quantum node/layer? (3) How can we model a classical counterpart to benchmark HQSL's classification performance? (4) Would HQSL scale as well as SL with an increasing number of clients?

\noindent
\textit{RQ 2:  How do we strengthen such hybrid quantum split learning schemes against data privacy leakage and reconstruction attacks? -- addressed in Section \ref{sec:reconstruction}}. 

Data privacy leakage from intermediate data transfers between clients and a server is an overarching problem in SL, as highlighted in Section \ref{sec:Literature Review}. It leads to private input data reconstruction attacks. In this work, we have the following sub-questions: (1) How do we model a threat scenario in which we can investigate HQSL's vulnerability to data privacy leakage and risks of reconstruction attacks? (2) Can we leverage the hybrid structure of HQSL to develop a countermeasure against reconstruction attacks? (3) How does this defense setup perform in the classical setting?

\section{\label{sec: HQSL} Hybrid Quantum Split Learning (HQSL)}
This section addresses our first research question, \textit{RQ 1}, as identified in Section \ref{sec:Problem_statement}. We first discuss some preliminaries around gate-based quantum circuits and their components necessary to understand our implementation of HQSL. Then, we describe how we design our HQSL model architecture. We discuss the construction and selection process for the quantum circuit introduced as a quantum layer into the server-side model of HQSL. We also describe how we devise the classical counterpart of the quantum node to allow benchmarking of HQSL's performance. We then propose 2 variants of HQSL depending on the data type they operate on and explain the training algorithm. Lastly, we describe an algorithm to scale HQSL to accommodate multiple clients. 
\subsection{Gate-based Quantum Circuits}
Before moving into the details of our HQSL model architecture, we first discuss the concept of gate-based quantum circuits, which are the fundamental components of quantum computing. Further background information on quantum computing can be found in Appendix \ref{sec: appendix-1}.  Gate-based quantum circuits (or quantum circuits) use quantum gates to manipulate qubits and perform computations during circuit evolution. Encoding classical data into quantum states is a critical initial step, achieved through various encoding methods such as basis, amplitude, or phase encoding. We can describe the encoding process generically as follows: $\phi(\vec{X}) = U_e(\vec{X})|0\rangle^{\otimes Q}$, where $\vec{X}$ is the classical data vector to encode, $|0\rangle^{\otimes Q}$ represents the initial quantum state of $Q$ qubits, typically all initialized to $|0\rangle$, and $U_e$ represents the unitary transformation that encodes the classical vector $\vec{X}$ into a quantum state. 
Another critical part of a quantum circuit is called the \textit{ansatz}, which is a predefined structure of quantum gates. The \textit{ansatz} is parameterized by a set of $k$ free parameters $\Theta=\{\theta_1, \theta_2, \dots, \theta_k\}$ that are optimized during training, e.g., in QML applications. We can represent the quantum state, $\Psi(\vec{X}, \vec{\Theta})$ at the end of the \textit{ansatz} as follows: $\Psi(\vec{X}, \vec{\Theta}) = U_p(\vec{\Theta})U_e(\vec{X})|0\rangle^{\otimes Q}$, where $U_p$ is the unitary representing the \textit{ansatz}. 
Quantum measurement, the final step of the circuit, involves collapsing the quantum state into a classical outcome, providing the results of computation. We measure a certain observable, $\hat{A}$, after each evolution of a quantum circuit and find the expectation value, E, of the measurement as 
\begin{equation}
    \text{E} = \langle\hat{A}\rangle = \langle\Psi|\hat{A}|\Psi\rangle \label{eq: measurementExp}
\end{equation} 
In our work, we use the Pauli-Y observable, i.e., $\hat{A}=\sigma_y$, where a projective measurement at each qubit would generate $\pm1$ which are the eigenvalues for the  $\sigma_y$ operator, for the qubit in the $\psi_{y_+}=\frac{1}{\sqrt{2}}\begin{bmatrix}1\\i\end{bmatrix}$ and $\psi_{y_-}=\frac{1}{\sqrt{2}}\begin{bmatrix}1\\-i\end{bmatrix}$ states respectively. The experimental implementation of measurements of a Q-qubit quantum circuit consists of evolving the quantum algorithm $M$ times (or shots) and computing the average for each qubit, $q \in \{1, \ldots, Q\}$, as an approximation of Eq. \ref{eq: measurementExp}, decomposed for each $q$ as,
\begin{equation}
    e_q = \langle\hat{A}\rangle_q \approx \frac{M_{-1,q}}{M}\cdot(-1) + \frac{M_{+1,q}}{M}\cdot(+1) \label{eq: approxmeasurementExp}, 
\end{equation}
where $M_{-1,q}$ and $M_{+1,q}$ are the number of times -1  and +1 measurements are obtained for qubit $q$, respectively. This average serves as an approximation of the probability that a given qubit is in a particular quantum state. For a Q-qubit system, we compute Eq. \ref{eq: approxmeasurementExp} for each $q  \in \{1, \ldots, Q\}$, and obtain a feature vector, $\vec{E}$, consisting of classical expectation measurement output at each qubit $q$ as,
\begin{equation}
    \vec{E} = (e_1, e_2, \ldots, e_Q). \label{eq: approxExpVec}
\end{equation}

The three steps described above (encoding of classical data into quantum states, parameterization via an ansatz, and measurement) form the foundation of Parameterized (or Variational) Quantum Circuits (PQC or VQC). Recent works have explored PQCs in various advanced configurations and structures, such as Block-Ring topologies for enhanced expressivity and entangling capacity \citep{liu_block-ring_2023}, and quantum convolutional structures for classification and code recognition tasks \citep{wu_degressive_2024}. However, for its practicality in the NISQ generation, this work investigates a small-scale quantum circuit featuring a novel data-loading technique, illustrated in Fig. \ref{fig:qlayer} and described in detail in the next section.
\subsection{Hybrid Quantum Split Learning Model}
\begin{figure}[hbt]
    \centering
    \includegraphics[]{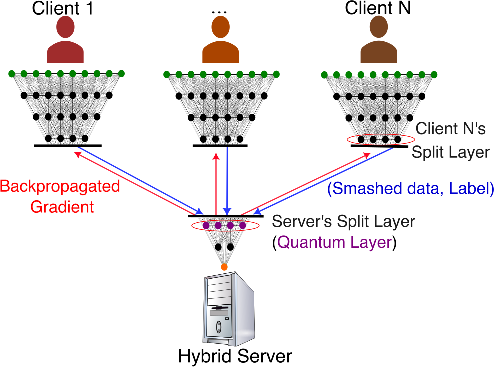}
    \caption{General structure of a Hybrid Quantum Split Learning (HQSL) model with N clients and 1 hybrid quantum server. The clients have a classical model (assumed to be the same for all clients), and the server model has a quantum layer as its first layer (layer with \textcolor{violet}{purple} units). The black, green, and orange units can be replaced by any classical neural network component (e.g., convolutional layers) to suit specific classification problems} 
    \label{fig:split_flow}
\end{figure}

While several SL configurations have been proposed \citep{vepakomma_split_2018}, this work considers the fundamental vanilla SL setup. In this setup, the labels and smashed data are transmitted to the server-side model at the split layer, while the raw data remain on the client side. During backpropagation, the gradients are transmitted from the server side to the client side across the split layer (see Fig. \ref{fig:split_flow}). In HQSL, the quantum layer is introduced as the first layer of the server-side model of SL. The quantum layer consists of a quantum circuit with classical features as input and output. The classical inputs are encoded to quantum states and processed by the quantum circuit. Expectation measurements of the resultant quantum states for each qubit lead to classical features output from the quantum layer (Eq. \ref{eq: approxExpVec}).
These are then fed to the next classical layer on the server-side model.
The quantum circuit is designed to be of practical importance in the near-term quantum computing era. A general structure of HQSL is shown in Fig. \ref{fig:split_flow}. Next, we discuss the construction and selection of the quantum circuit used in this work.

\subsubsection{\label{subsubsec: qlayer} Construction and Selection of Quantum Circuit}
\begin{figure}[htbp!]
\centering
    \begin{subfigure}{0.3\textwidth}
        \centering
        \includegraphics[width=\textwidth]{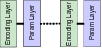}
        \caption{Data-loading scheme}
        \label{fig:schem}
    \end{subfigure}
    \begin{subfigure}{0.6\textwidth}
        \centering
        \includegraphics[width=\textwidth]{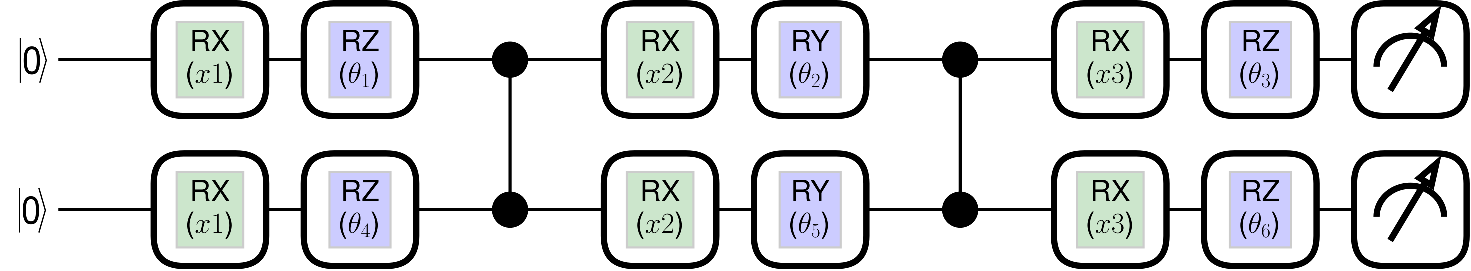}
        \caption{Circuit 6: Proposed quantum circuit}
        \label{fig:circuit}
    \end{subfigure}
    \caption{(a) Our qubit-efficient data-loading scheme consists of alternate layers of encoding and parameterized (Param) gates. (b) The quantum circuit that we utilize in this work has RX gates serving as data-loading points (encoding layer). It is designed to make efficient use of qubits while loading data to the quantum circuit}
    \label{fig:qlayer}
\end{figure}

To construct the quantum circuit in HQSL's quantum layer, we consider the limitations of current-generation quantum computers (see Appendix \ref{sec: appendix-1} for details).
Specifically, the circuit width (number of qubits) and depth (maximum number of gates per qubit) are critical considerations that need to be kept as small as possible. The presence or absence of entangling gates is also another important consideration. 
We trialed 6 different quantum circuit configurations by adjusting the number of qubits, entangling gates, and encoding type.  Each configuration was tested by incorporating the circuit as a quantum layer in our HQSL model. The quantum circuit was selected based on the comparison of accuracy and F1-score between HQSL and its classical analogue. We chose the circuit that consistently outperformed SL in all our experiments. We detail these experiments in Appendix \ref{subsec:appendix1}.

The quantum circuit with the optimal performance is shown as Circuit 6 in Fig. \ref{fig:circuit}. The circuit consists of two qubits, each receiving a 3-dimensional classical input feature vector $\vec{X}=(x_1, x_2, x_3)$. These features are encoded into each qubit at three distinct points along the circuit using RX-gates. This encoding strategy is inspired by the data re-uploading technique proposed by \citep{perez-salinas_data_2020}, where classical information is introduced multiple times (using U-gates) throughout the circuit layers to enhance representational capacity. 
However, our designed circuit differs in two fundamental ways by: (i) encoding input data $\vec{X}\in\mathbb{R}^n$ at $n$ loading points ($n=3$ for our case) using single-qubit RX-rotations alternating between single-qubit parameterized gates without using U-gates, and (ii) avoiding layer repetitions (re-uploading) to achieve lower circuit depth and fewer parameters while improving performance (see Table \ref{tab:circuits_678}). We provide more details on these differences and how we designed our circuit next.

The data re-uploading technique involves multiple repetitions of a layer, $L_i$, consisting of the generic single-qubit rotation gate, $U \in SU(2)$. The U-gate is used in the encoding layer as  $U(\vec{X})$ and in the parameterized layer as $U(\vec{\Theta})$, where $\vec{X} = (x_1, x_2, x_3)$ and $\vec{\Theta_i} = (\theta_1^i, \theta_2^i, \theta_3^i)$, where $\vec{\Theta_i}$ represents the parameters for layer $L_i$. Thus, each layer can be represented as $L_i = U(\vec{X}) U(\vec{\Theta_i})$. $L_i$, for $i=1, \ldots, N$, is repeated, giving rise to the data re-uploading mechanism, and the single-qubit classifier is introduced, given enough repeats, $N$, are performed.

The U-gate consists of parameters $\vec{\Phi} = (\phi_1, \phi_2, \phi_3)$ and can be expressed as $U(\phi_1, \phi_2, \phi_3) \in SU(2)$. We considered the following decomposition of the U-gate: 
\begin{equation}
    U(\vec{\Phi}) = U(\phi_1,\phi_2,\phi_3) = RZ(\phi_3) RY(\phi_2) RZ(\phi_1)
\end{equation}

\begin{figure}
    \centering
    \includegraphics[width=0.8\linewidth]{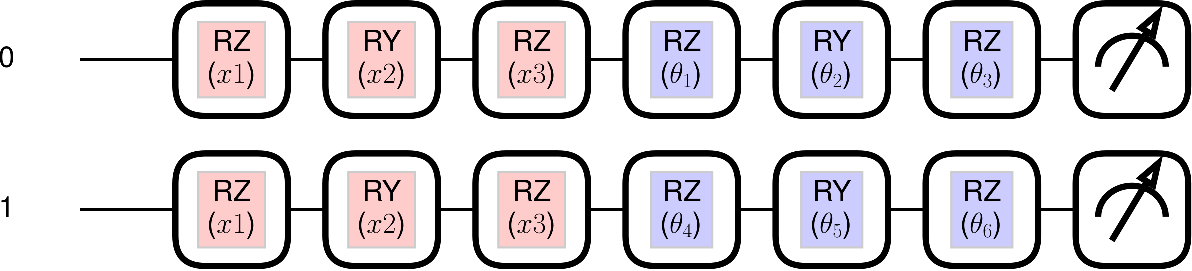}
    \caption{Circuit 7: Representation of a 2-qubit circuit with a single layer, $L$, without entanglement consisting of the arrangement $U(\vec{X}) - U(\vec{\Theta})$, where the U-gate is decomposed here in RZ--RY--RZ gates for each qubit}
    \label{fig:orig_dru}
\end{figure}

\begin{figure}
    \centering
    \includegraphics[width=0.9\linewidth]{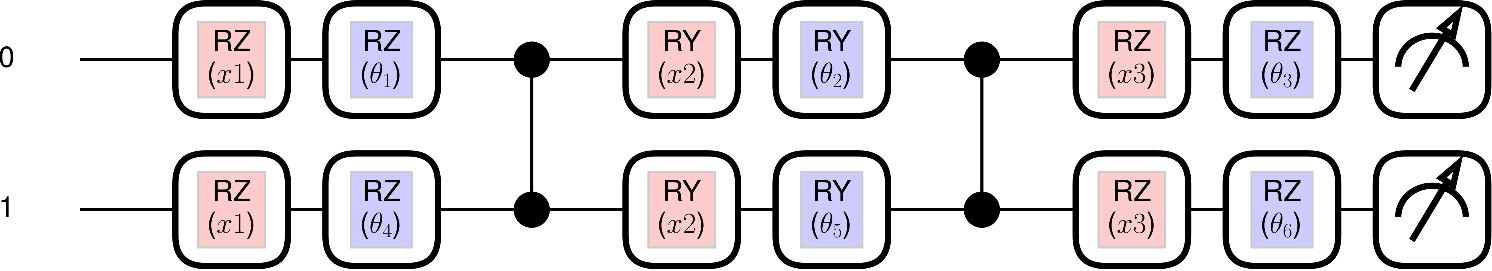}
    \caption{Circuit 8: Representation of a reorganization of Circuit 7 shown in Fig. \ref{fig:orig_dru} to introduce CZ entanglement between each qubit}
    \label{fig:orig_dru_with_ent}
\end{figure}

For our work, we adapted the above decomposition of this layer, $L_i$, taking into consideration that the width and depth of a circuit should be kept as small as possible, given the limitations due to current quantum devices. We considered a circuit consisting of only a single layer $L$ consisting of 2 qubits, and without any entangling gates. This circuit has a depth of 6 and consists of a total of 6 trainable parameters ($\theta_1, \ldots, \theta_6$). We represent this circuit as Circuit 7 in Fig. \ref{fig:orig_dru}.
As discussed by \citep{perez-salinas_data_2020}, entangling gates can improve the classification performance of the circuit. To introduce entanglement within the circuit consisting of a single layer, $L_i$, we decomposed the U-gate into RZ--RY--RZ gates and reorganized the circuit such that for each qubit, we uploaded a feature, $x_k$ for $k \in \{1,2,3\} $ using rotational gates, followed immediately by a single-qubit parameterized rotational gate of the same type, and a CZ entangling gate between the 2 qubits. This is repeated until all the features have been uploaded. We omit the CZ gate at the end of the circuit. Hence, this circuit still has a total of 6 trainable parameters, but has now a depth of 8. We show this circuit as Circuit 8 in Fig. \ref{fig:orig_dru_with_ent}.

We built both circuits 7 and 8 and tested their performance when utilized in the quantum layer of HQSL. We compared their performance for the classification of the Fashion-MNIST dataset \citep{xiao2017fashionmnist} using the accuracy and F1-score metrics--we experimented only on the Fashion-MNIST dataset because it had more room for improvement compared to the other datasets we studied in this paper. All the details for this dataset and these experiments are provided in Section \ref{sec:hqsl_exp}. The results, shown in Table \ref{tab:circuits_678}, demonstrate that the use of entanglement indeed can improve the accuracy and F1-score during the classification of the Fashion-MNIST dataset at the expense of 2 additional CZ gates, increasing the depth by 2, but keeping the number of trainable parameters to 6.

To develop our proposed circuit (Circuit 6), we carried out a final modification to Circuit 8. Specifically, we replaced the embedding gates for loading $\vec{X} = (x_1,x_2,x_3)$ into the circuit with RX-gates, as shown in Fig. \ref{fig:circuit}. We note that this modification kept the number of trainable parameters and circuit depth of our proposed circuit similar to Circuit 8.
This change to RX gates for encoding further improved the classification performance, as shown in Table \ref{tab:circuits_678}.
\begin{table}[htbp!]
    \caption{Table comparing the quantum resource utilization (Number of Trainable Parameters, and Circuit Depth) and 5-fold Mean Accuracy and F1-score of HQSL models consisting of Circuits 7, 8, 6, 9, 10 during classification of Fashion-MNIST dataset. Circuits 9 and 10 are 3-layer versions of Circuit 7 ($L_i(j)$), demonstrating the data re-uploading technique with and without entangling CZ gates, respectively ($L_i(j)$ represents the $j^{th}$ layer on qubit $i$). Circuit 6, which is a modified version of Circuits 7 and 8, achieves $1-2\%$ accuracy and F1-score improvements with fewer parameters and more than half the circuit depth of standard data re-uploading circuits 9 and 10}
    \centering
    \begin{tabular}{cccccc}
    \hline
        ID & Circuit & Number of Trainable & Circuit & Mean  & Mean \\
         &  &Parameters& Depth &Accuracy &  F1-score
        \\
        \hline
         7 &\begin{minipage}{.3\textwidth}
        \vspace{2ex}
      \includegraphics[width=\linewidth]{Fig4.eps}
    \end{minipage} \vspace{1ex} & 6&6&0.812 &0.815\\
    \hline
         8 & \begin{minipage}{.3\textwidth}
        \vspace{2ex}
      \includegraphics[width=\linewidth]{Fig5.eps}
    \end{minipage} \vspace{1ex} &6&8&0.823&0.825\\
    \hline
         6 & \begin{minipage}{.3\textwidth}
        \vspace{2ex}
      \includegraphics[width=\linewidth]{Fig3.eps}
    \end{minipage} \vspace{1ex} & 6&8&\textbf{0.839}&\textbf{0.840}\\
\hline
         9 & \begin{minipage}{.3\textwidth}
        \vspace{2ex}
      \includegraphics[width=\linewidth]{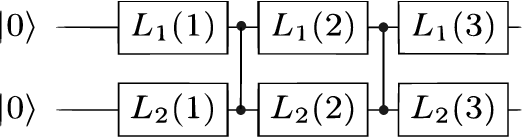}
    \end{minipage} \vspace{1ex} &18&20&0.824&0.826\\
\hline
        10 & \begin{minipage}{.3\textwidth}
        \vspace{2ex}
      \includegraphics[width=\linewidth]{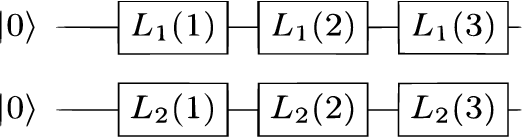}
    \end{minipage} \vspace{1ex}  
 &18&18&0.821&0.823\\
\hline
    \end{tabular}  
     \label{tab:circuits_678}
\end{table}
The last two rows of Table \ref{tab:circuits_678} show the performance when we adopt the standard data re-uploading technique in a 2-qubit circuit with 3 repeats of layer $L_i$, with (Circuit 9) and without entanglement (Circuit 10), respectively. We used the same entanglement strategy proposed in \citep{perez-salinas_data_2020}, i.e., using CZ-gates after each layer $L_i$, except the last layer. In Table \ref{tab:circuits_678}, we also provide a comparison of quantum resource utilization in terms of the number of trainable parameters, and circuit depth of our proposed quantum circuit compared to the data re-uploading circuits proposed in \citep{perez-salinas_data_2020}. 

Table \ref{tab:circuits_678} shows that our proposed shallow (8-deep with 6 trainable parameters) qubit-efficient data-loading circuit (Circuit 6) provides better accuracy and F1-score than the much deeper circuits (Circuits 9 and 10) that use the standard data re-uploading scheme. By loading features exclusively with one-type of single-qubit gates (RX-gates in our case) and omitting repeated layers, Circuit 6 maintains a shallow depth (8) and minimal trainable parameter count (6), compared to the 3-layer data re-uploading circuits (Circuits 9 \& 10), which require 18 parameters and depths of 18–20. Our proposed data loading method allows us to achieve better accuracy and F1-score with fewer parameters, improving the trainability and efficiency of our design compared to the standard data re-uploading circuits.
\textit{Our proposed circuit 6 is `qubit-efficient', as it provides a method to encode a classical input feature vector, $\vec{X}$, of arbitrary dimension onto the same qubit. However, this comes at the cost of increasing circuit depth proportionally with input dimensionality.}
Hence, in this work, we restrict the circuit depth by only considering 3-dimensional data being loaded onto a 2-qubit quantum circuit. \textit{By imposing this restriction, we also keep our simulations within reasonable runtimes, while still providing high performance during classification, with a shallow quantum circuit.}

\begin{figure}[htbp!]
    \centering
    \includegraphics[width=0.9\textwidth, height=!]{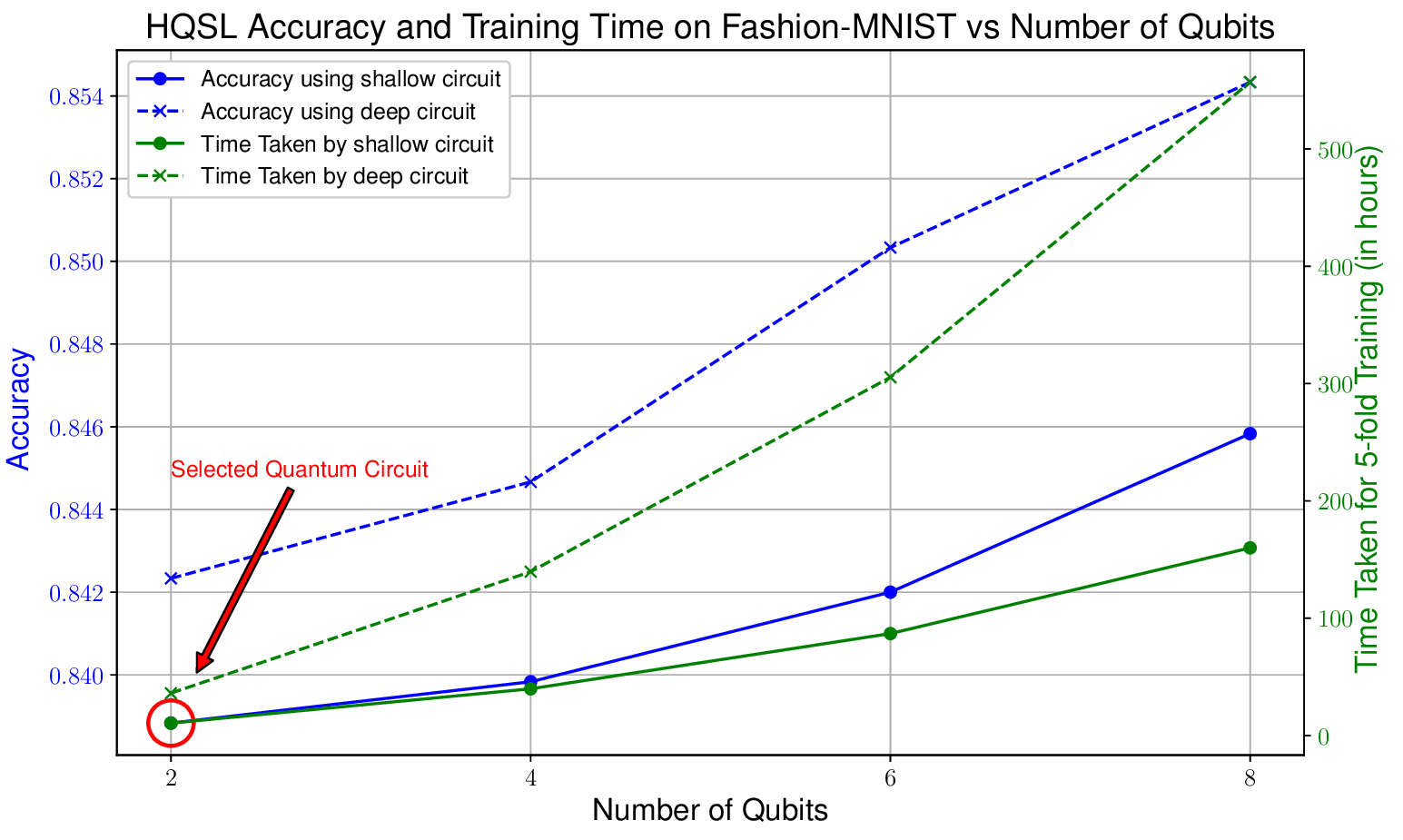}
    \caption{Effect of the number of qubits and circuit depth on the mean test accuracy and 5-fold training time of HQSL on Fashion-MNIST dataset. Accuracy improves only slightly as we increase the number of qubits and circuit depth at the expense of rapidly increasing training time}
    \label{fig:fmnist_vs_n_qubits}
\end{figure}

To demonstrate this, we carried out experiments with Circuit 6 by increasing the number of qubits and circuit depth. The resulting accuracy, and simulation time taken for 5-fold training can be seen in Fig \ref{fig:fmnist_vs_n_qubits}. 
In the figure, `shallow' refers to a low-depth circuit with 3 data loading points, while `deep' corresponds to a circuit twice as deep as the shallow one,  i.e., 6 data-loading points. Hence, a shallow circuit would require 3-dimensional classical features, while a deep circuit would require a data dimension of 6.

We highlight here that although as we increased the number of qubits and circuit depth, accuracy increased marginally, the runtime during simulation increased very rapidly (comparing the shallow circuit with 2 qubits and the deep circuit with 8 qubits, an improvement of only around 1.5\% causes the time taken for training to increase by more than 500 hours).
We also highlight that the deeper (maximum number of gates) and wider (number of qubits) a circuit is, the more difficult it is to implement on currently available quantum computers. We also face the risk of the barren plateau phenomenon when the number of trainable parameters is excessively high resulting in vanishing gradients \citep{mcclean2018barren}.
For these reasons, it is desirable to keep our quantum circuit shallow and narrow (3 data-loading points and 2 qubits), while still obtaining high accuracy. In Section \ref{subsec: exp_1client_tests}, we effectively provide initial empirical evidence that HQSL with our proposed quantum circuit can be deployed on currently available quantum hardware, providing performance similar to noise-free simulations.
\textit{Hence, our empirical findings show that our proposed quantum circuit keeps the training time during simulations within reasonable limits, is more likely to be feasible on near-term devices, and provides a better classification performance than a deep circuit with the data re-uploading technique comprising of 3 repeats of layer $L_i$.} 

This qubit-efficient data-loading methodology for quantum circuit design can be extended beyond the scope of HQSL, presenting a versatile framework with broad applicability. Further research is warranted to comprehensively evaluate its wider implications and potential use cases across various domains, which is beyond the scope of this paper.
%
\subsubsection*{Classical Benchmarking of HQSL}
Our objective is to create an equivalent classical model to benchmark the performance of HQSL. To achieve this, we construct the classical counterpart of our proposed quantum circuit (Circuit 6) as follows. It consists of a dense layer with the same number of output neurons as the number of qubits (or measurement outputs) of the quantum layer -- two output neurons with \code{ReLU} activation units (the best-performing activation unit based on our experiments). The number of input neurons in this layer corresponds to the dimension of the classical input (3 input features). 
The aim is to compare our compact quantum layer to a simple classical dense layer, enabling a fair comparison, taking into consideration the size limitations of quantum circuits. In Appendix \ref{subsec:appendix1}, we provide a comparative analysis of the performances of different quantum circuits and their corresponding classical counterparts when introduced as a layer in the split learning models. This experimental analysis allowed us to construct and select the quantum circuit to be used in HQSL, as we described earlier.
\subsubsection{\label{subsubsec:HQSL_modelling} HQSL Model Variants} 
\begin{figure}[htpb]
   \begin{subfigure}{\textwidth}
    \centering
        \includegraphics[]{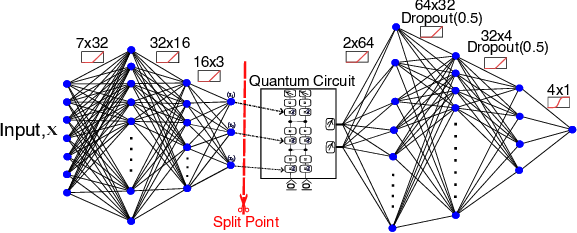}
        \caption{HQSL variant 1}
        \label{fig:model1}
    \end{subfigure}
    \medskip
    \begin{subfigure}{\textwidth}
    \centering
        \includegraphics[]{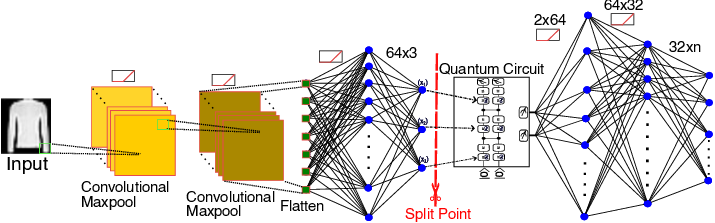}
        \caption{HQSL variant 2}
        \label{fig:model2}
    \end{subfigure}
    \caption{HQSL variants: The split point splits the network into the client-side portion (left) and server-side portion (right). (a) HQSL variant 1: For binary classification on multivariate datasets. (b) HQSL variant 2: For $n$-class classification on single-channel image datasets. We can set the number of output nodes, $n$, depending on the number of classes in the dataset}
    \label{fig:HQSLmodels}
\end{figure}

We propose two variants of HQSL designed for multivariate binary and multi-class single-channel image dataset classification. Our approach involves constructing an HQNN and subsequently dividing it at the \textit{split point} (see Fig. \ref{fig:HQSLmodels}) so that the initial part of the HQNN represents the client side (classical neural network layers) and the remaining segment functions as the server side (hybrid quantum neural network layers). This explicit division or splitting of the HQNN into client-side classical layers and server-side hybrid quantum-classical layers constitutes the key novelty of HQSL, distinguishing it from existing HQNN approaches that do not address resource-constrained deployment or collaborative training frameworks.

We assume here that the clients, e.g., IoT devices, have sufficient classical resources to perform client-side computations and lack quantum resources, as stated in Section \ref{sec:Problem_statement}. This is a fair assumption because typical IoT devices have sufficient processing capabilities to execute lightweight classical neural network models, but generally lack the resources and infrastructure for quantum computations. The client-side model portion extracts compressed feature representations (smashed data), to transmit low-dimensional data to the server-side model portion, reducing communication overhead and enabling a shallow quantum circuit in the quantum layer for further processing on the server side.

\noindent
\textbf{HQSL variant 1:} The client-side model consists of 4 fully connected dense layers with \code{ReLU} activation functions with a 7-dimensional input and outputs smashed data of dimension 3. For the server-side model, we start with the quantum layer, constructed using our proposed quantum circuit in Section \ref{subsubsec: qlayer}, which has an input dimension of 3 and an output dimension of 2. We concatenate 4 additional fully connected classical dense layers with the \code{ReLU} activation function (except the last layer, which uses \code{Sigmoid}) with an output dimension of 1 (for binary classification). We apply \code{Dropout} layers on the first two classical layers on the server-side model to improve regularization. 
For the classical counterpart of this variant, we replace the quantum layer with the classical dense layer as described earlier.

\noindent
\textbf{HQSL variant 2:} The client-side model consists of two Convolutional Neural Network filters, each with \code{ReLU} activation function and a \code{MaxPool2d} Layer, with a stride of 2 and kernel size of 2, followed by a \code{flatten} layer, and 2 fully connected layers to bring the feature dimensions down to 3. This is the client-side portion of the model.  
For the server side, we introduce our quantum circuit as a quantum layer and concatenate it with 3 more fully connected classical dense layers with an output dimension of $n$ (for $n$-class classification). We apply the \code{ReLU} activation function on the first 2 classical layers on the server side. 
For the classical version, we replace the quantum layer with its equivalent classical dense layer, acting as a benchmark for this HQSL variant.
Fig. \ref{fig:HQSLmodels} depicts the two HQSL model variants that we study in this work. 

\subsubsection{\label{subsubsec:opt}Hybrid Quantum Split Learning Model Training}
In our HQSL framework, the training process follows a client-server paradigm where data is processed in two distinct stages. The client side, which consists of a classical neural network, processes the raw input data $\vec{X}$ and extracts feature representations. These extracted features, referred to as the smashed data $\vec{Z}$, are then transmitted to the server side, together with the true labels, $\vec{Y}$.

The server side consists of a hybrid quantum-classical sub-model, where, first, the quantum layer processes $\vec{Z}$, and outputs measurement expectation values $\vec{E}$. $\vec{E}$ is now input to the remaining layers of the server-side model portion which, in our HQSL model, are all classical dense layers with $\hat{\vec{Y}}$ as the final output. This structured split ensures that the client operates solely with classical resources, while the server-side quantum component allows the client to benefit from quantum processing without requiring direct access to quantum hardware. The placement of the quantum layer as the first layer of the server-side model portion is a deliberate design choice aimed at enhancing resistance to reconstruction attacks in split learning, as further discussed in Section \ref{sec:reconstruction}.
Next, we outline the computation and backpropagation of gradients through the classical and quantum layers of HQSL to enable effective training.

For the classical layers of HQSL, gradients are computed from the loss function, $\mathcal{L(\vec{Y}, \hat{Y})}$, using the classical backpropagation algorithm via the chain rule. 
However, it becomes non-trivial to do the same for the quantum layer. Schuld et al. devised the parameter-shift rule, which permits backpropagation through the quantum layer \citep{schuld_evaluating_2019}. We briefly discuss the main ideas in the following.

Consider a quantum circuit, $f_q: \vec{Z} \mapsto \vec{E}$ ($f_q: \mathbb{R}^{M} \rightarrow \mathbb{R}^{N}$), where $f_q$ is parameterized by a set of $k$ free parameters, $\vec{\Theta} = \{\theta_1, \theta_2, \ldots, \theta_k\}$, and maps input, $\vec{Z}=(z_1, z_2, \ldots, z_{M})$ to the measurement outputs, $\vec{E}=(e_1, e_2, \ldots, e_{N})$. In HQSL, the quantum layer can be considered as being sandwiched between two sets of classical layers: the client-side model layers, $f_c: \vec{X} \mapsto \vec{Z}$ $(f_c:\mathbb{R}^{in}\rightarrow \mathbb{R}^{M})$, and the server-side classical layers, $f_s: \vec{E} \mapsto \hat{Y}$ $(f_s:\mathbb{R}^{N}\rightarrow \mathbb{R}^{out})$, where $\hat{Y}$ is the output of HQSL. The smashed data $\vec{Z}$ is then transferred to the server side to continue forward propagation, together with the true labels $\vec{Y}$, to compute the loss $\mathcal{L}$.

The derivatives of the expectation value of the measurement, $\vec{E}$, of the quantum circuit, $f_q$ with respect to the gate parameters, $\vec{\Theta}$, are computed by evolving the circuit twice per parameter, with a ($\pm$) shift in that parameter. 
Representing the partial derivative of $f_q(\vec{\Theta}, \vec{Z})=\vec{E}$ with respect to parameter $\theta_i$ for $i$ $\in$ $\{1, 2, \ldots, k\}$ as $\frac{\delta{f_q}}{\delta{\theta_i}}$, we can express the parameter-shift rule as:
\begin{equation}
    \frac{\delta{f_q}}{\delta{\theta_i}} = r[f_q(\theta_i+\frac{\pi}{4r}, \vec{Z}) + f_q(\theta_i-\frac{\pi}{4r}, \vec{Z})] = \frac{\delta{\vec{E}}}{\delta{\theta_i}}, \label{eq:paramshift}
\end{equation}
where $r$ is a shift constant.
This gives us the exact gradient with respect to each free parameter $\theta_i$ in the quantum circuit. Similarly, we can compute the derivative with respect to input $z_j$, for $j \in \{1, 2, \ldots, M\}$, to the quantum circuit:
\begin{equation}
    \frac{\delta{f_q}}{\delta{z_j}} = r[f_q(\vec{\Theta}, z_j+\frac{\pi}{4r}) + f_q(\vec{\Theta}, z_j-\frac{\pi}{4r})] = \frac{\delta{\vec{E}}}{\delta{z_j}} \label{eq:paramshiftInput}
\end{equation}

In summary, on the server side, for the quantum layer, during forward propagation, the smashed data $\vec{Z}$ are fed as classical features to the quantum layer, and the quantum circuit is evolved for a default of 1000 shots. At the end of the quantum circuit evolutions, an expectation value for the measurement is calculated as per Eq. \ref{eq: approxmeasurementExp}. These calculated values are then used as classical features for the next classical layer. 
During backpropagation, to compute the derivatives with respect to parameter $\theta_i \in \{\theta_1, \theta_2, \ldots, \theta_k\}$ and inputs $z_i \in \{z_1, z_2, \ldots, z_M\}$, two evaluations of the circuit are computed with a $\pm$ shift in the parameter $\theta_i$. These two computations are then added according to Eq. \ref{eq:paramshift} and Eq. \ref{eq:paramshiftInput}. This is repeated until the derivatives with respect to all $\theta_i$ and $z_j$ are computed. Using the computed derivatives, we update the gate parameters, $\vec{\Theta}$, accordingly. The derivative $\frac{\delta{f_q}}{\delta{z_j}}$ is used to compute the gradient with respect to the quantum layer input $z_j$ as $\frac{\delta{\mathcal{L}}}{\delta{z_j}}$ using the chain rule. This gradient is transmitted across the split layer to the client-side model to continue the backpropagation of the loss function, $\mathcal{L}$, to the client-side classical layers. 
We summarize the HQSL training in Algorithm \ref{alg:hybsplit}. In the algorithm, $\frac{\delta{(\cdot)}}{\delta{(\cdot)}}$ represents the Jacobian matrix that stores the element-wise gradients of the vector in the numerator with respect to the vector in the denominator.
\begin{algorithm}[htbp!]

    \caption{Hybrid Quantum Split Learning Training with a Single Classical Client and Hybrid Quantum Server}
    \label{alg:hybsplit}
    \textbf{Input: }Number of Training epochs $N$, learning rate $\eta$ 
    
    \textbf{Output: }Optimal weights, $\vec{\Tilde{W}} = (\vec{\Tilde{W}}^C, \vec{\Tilde{W}}^S)$, for classical layers and optimal parameters, $\vec{\Tilde{\Theta}}$, for quantum layer

    \textbf{Notations: }Client's inputs: $\vec{X}$, client's labels: $\vec{Y}$, predicted labels: $\hat{Y}$, client's smashed data: $\vec{Z}$, quantum layer expectation measurement output: $\vec{E}$, loss function: $\mathcal{L}$

    \textbf{Initialization: } Initialize client-side layers ($f_c$) weights, $\vec{W}^C$; server-side quantum layer ($f_q$) parameters, $\vec{\Theta}$, and classical layers ($f_s$) weights, $\vec{W}^S$.

    \medskip
    \textbf{START OF TRAINING}
    
    \For{$n={1, \dots, N}$ Training Epochs}{
            \textbf{Client side:} \tcp{Forward Pass}
            
                \Indp$\vec{Z} = f_c(\vec{X},\vec{W}_n^C)$; Send $\vec{Z}$ and $\vec{Y}$ to the server
                
                \Indm\dotfill{}
                
             \textbf{Server side:} \tcp{Forward Pass, Compute Loss, Backward Pass}

                    \Indp\textit{Quantum Layer:} 
                    $\vec{E} = f_q(\vec{Z},\vec{\Theta}_n)$ \tcp{Compute $\vec{E}$ as per Eq. \ref{eq: approxExpVec}}
                                    
                    \textit{Classical Layers:}
                    $\hat{Y} = f_s(\vec{E},\vec{W}^S_n)$
                    
                    \Indp  Compute Loss, $\mathcal{L}=\mathcal{L}(\vec{Y},\hat{Y})$, and gradient w.r.t. output $\hat{Y}$ as $\frac{\delta{\mathcal{L}}}{\delta{\hat{Y}}}$ 

                    Compute $\frac{\delta{\hat{Y}}}{\delta{\vec{W}^S_n}}=\frac{\delta{f_s(\vec{E},\vec{W}^S_n)}}{\delta{\vec{W}^S_n}}$ and $\frac{\delta{\hat{Y}}}{\delta{\vec{E}}} = \frac{\delta{f_s(\vec{E},\vec{W}^S_n)}}{\delta{\vec{E}}}$

                    Compute gradient w.r.t. $\vec{W}^S_n$ as $\frac{\delta{\mathcal{L}}}{\delta{\vec{W}^S_n}} = \frac{\delta{\mathcal{L}}}{\delta{\hat{Y}}} \cdot \frac{\delta{\hat{Y}}}{\delta{\vec{W}^S_n}}$ \tcp{Chain rule}

                    Compute gradient w.r.t. $\vec{E}$ as: $\frac{\delta{\mathcal{L}}}{\delta{\vec{E}}} = \frac{\delta{\mathcal{L}}}{\delta{\hat{Y}}} \cdot \frac{\delta{\hat{Y}}}{\delta{\vec{E}}}$ \tcp{Chain rule} 

                    Update server classical layers weights $\vec{W}^S$ as: $\vec{W}^S_{n+1} \leftarrow \vec{W}^S_n - \eta \frac{\delta{\mathcal{L}}}{\delta{\vec{W}^S_n}}$

                    \Indm\textit{Quantum Layer:}

                    \Indp Compute $\frac{\delta{\vec{E}}}{\delta{\vec{\Theta}_n}}=\frac{\delta{{f_q(\vec{Z},\vec{\Theta}_n)}}}{\delta{\vec{\Theta}_n}}$ and $\frac{\delta{\vec{E}}}{\delta{\vec{Z}}} = \frac{\delta{f_q(\vec{Z},\vec{\Theta}_n)}}{\delta{\vec{Z}}}$ \tcp{Parameter-shift rule Eq. \ref{eq:paramshift}, \ref{eq:paramshiftInput}}

                    Compute gradient w.r.t. $\vec{\Theta}_n$ as $\frac{\delta{\mathcal{L}}}{\delta{\vec{\Theta}_n}} = \frac{\delta{\mathcal{L}}}{\delta{\vec{E}}} \cdot \frac{\delta{\vec{E}}}{\delta{\vec{\Theta}_n}}$ \tcp{Chain rule}

                    Compute gradient w.r.t. $\vec{Z}$ as $\frac{\delta{\mathcal{L}}}{\delta{\vec{Z}}} = \frac{\delta{\mathcal{L}}}{\delta{\vec{E}}} \cdot \frac{\delta{\vec{E}}}{\delta{\vec{Z}}}$ \tcp{Chain rule}

                    Update Quantum Layer Parameters $\vec{\Theta}$ as: $\vec{\Theta}_{n+1} \leftarrow \vec{\Theta}_n - \eta \frac{\delta{\mathcal{L}}}{\delta{\vec{\Theta}_n}}$

                    Send $\frac{\delta{\mathcal{L}}}{\delta{\vec{Z}}}$ to client side to continue backward pass

            \Indm \Indm \dotfill{}
            
            \textbf{Client side: }\tcp{Backward Pass}
            
                \Indp Compute $\frac{\delta{\vec{Z}}}{\delta{\vec{W}^C_n}} = \frac{\delta{f_c(\vec{X},\vec{W}_n^C)}}{\delta{\vec{W}_n^C}}$ 

                Compute gradient w.r.t. $\vec{W}^C_n$ as 
                $\frac{\delta{\mathcal{L}}}{\delta{\vec{W}^C_n}} =  \frac{\delta{\mathcal{L}}}{\delta{\vec{Z}}} \cdot \frac{\delta{\vec{Z}}}{\delta{\vec{W}_n^C}}$ \tcp{Chain rule} 

                Update client model weights $\vec{W}^C$ as: 
                $\vec{W}^C_{n+1} {\leftarrow} \vec{W}^C_n - \eta \frac{\delta{\mathcal{L}}}{\delta{\vec{W}^C_n}}$             
                    
            \medskip 
            \Indm \textbf{End of Epoch $n$}
            }            
        \medskip 
    \textbf{END OF TRAINING}
    
    \Return $\vec{\tilde{W}}{=}(\vec{\tilde{W}}^C, \vec{\Tilde{W}}^S$), $\vec{\tilde{\Theta}}$
\end{algorithm}

\subsection{\label{subsubsec:scalability}Hybrid Quantum Split Learning with Multiple Clients}

Scaling HQSL to accommodate a larger number of clients enables multiple clients to leverage the quantum resources at the central server. The method we use to scale HQSL with $\textsf{K}$ clients follows the basic round-robin protocol and is as follows. We initialize the model parameters of the client side and server side as $\vec{W^C}$ and $(\vec{\Theta}, \vec{W}^S)$ respectively. $\vec{\Theta}$ corresponds to the quantum layer trainable gate parameters that lie on the server side. We randomly choose a client $\textsf{k}$ ($\textsf{k} \in \{0,1,2, \dotsc, \textsf{K}-1\}$) and train it in collaboration with the server. This consists of one round of forward and backward propagation and makes 1 local epoch for client $\textsf{k}$. The updated client $\textsf{k}$ model weights $\vec{W}^{C'}$ are then sent to the next client ($\textsf{k}'$) that updates its model weights before another round of forward and backward propagation with the server. This marks the end of client $\textsf{k}'$'s local epoch. After all $\textsf{K}$ clients have been served in that 1 global epoch, we move to the next global epoch and resume training client $\textsf{k}$. It is important to highlight that during each communication round between a client and the server, only low-dimensional data is transmitted, as the output layer of the client-side model consists of only 3 nodes, as shown in Fig. \ref{fig:HQSLmodels}. This significantly reduces the communication overhead, especially when considering a very large number of clients, e.g., IoT devices in edge networks. A detailed analysis or optimization of communication efficiency is beyond the scope of this work. We leave a thorough investigation of strategies to further minimize communication overhead in HQSL for future research.

The differences due to the presence of the quantum layer in HQSL are as follows. A forward and backward propagation round consists of encoding classical smashed data within the quantum layer. At the measurement stage, we use a default of 1000 shots to sample the expectation value of the measurements due to the probabilistic nature of quantum circuits. 
Finally, we use the parameter-shift rule described earlier for computing gradients with respect to the parameters ($\vec{\Theta}$) and inputs ($\vec{Z}_\textsf{k}$) of the quantum circuit. We summarize the method for including multiple clients in Algorithm \ref{alg:hybsplitN_Clients} (presented in two parts: Initialization and Training Phase).
\begin{algorithm}
    \caption{Hybrid Quantum Split Learning Training with $\textsf{K}$ Classical Clients and Hybrid Quantum Server (Initialization Phase)} 
    \label{alg:hybsplitN_Clients}
    \textbf{Input: } Number of clients $\textsf{K}$, number of training epochs $N$, learning rate $\eta$ 
    
    \textbf{Output: } Optimal weights, $\vec{\Tilde{W}}=(\vec{\Tilde{W}}^C, \vec{\Tilde{W}}^S)$, for classical layers and optimal parameters, $\vec{\Tilde{\Theta}}$, for quantum layer

    \textbf{Notations: } Client $\textsf{k}$'s inputs: $\vec{X}_{\textsf{k}}$, labels: $\vec{Y}_{\textsf{k}}$, predicted labels: $\hat{Y}_{\textsf{k}}$, smashed data: $\vec{Z}_{\textsf{k}}$, quantum layer outputs when client $\textsf{k}$ is being served: $\vec{E}_{\textsf{k}}$, loss function: $\mathcal{L}$

    \textbf{Initialization: } Initialize client-side layers ($f_c$) weights, $\vec{W}^C$; server-side quantum layer ($f_q$) parameters, $\vec{\Theta}$, and classical layers ($f_s$) weights, $\vec{W}^S$. 
\end{algorithm}
    \SetNlSty{texttt}{(}{)}
\begin{algorithm}
    \textbf{START OF TRAINING}
    
    \caption*{Hybrid Quantum Split Learning Training with $\textsf{K}$ Classical Clients and Hybrid Quantum Server (Training Phase)}
      
    \For{$n={1, \ldots, N}$ Training Epochs}{

            \textbf{Start of global epoch $n$}
            
            \For{each client \textup{$\textsf{k}$}, \textup{$\textsf{k} \in  \{0, \ldots, \textsf{K}-1$\}}}{
            
            Set next client $\textsf{k}^\prime$ as $\textsf{k}^\prime = (\textsf{k}+1)\pmod{\textsf{K}}$
            
            \textbf{Start of client $\textsf{k}$'s local epoch} 
            
            \textbf{Client $\textsf{k}$ side:} \tcp{Forward Pass}
            
            \Indp$\vec{Z}_{\textsf{k}} = f_c(\vec{X}_{\textsf{k}}, \vec{W}_{n,\textsf{k}}^C)$; Send $\vec{Z}_{\textsf{k}}$ and $\vec{Y}_{\textsf{k}}$ to server
                        
            \Indm \dotfill{}
            
            \textbf{Server side:} \tcp{Forward Pass, Compute Loss, Backward Pass}

                    \Indp\textit{Quantum Layer:} $\vec{E}_{\textsf{k}} = f_q(\vec{Z}_{\textsf{k}}, \vec{\Theta}_{n,\textsf{k}})$ \tcp{Compute $\vec{E}_\textsf{k}$ as per Eq. \ref{eq: approxExpVec}}
                    
                    \textit{Classical Layers:} $\hat{Y}_{\textsf{k}} = f_s(\vec{E}_{\textsf{k}},\vec{W}_{n,\textsf{k}}^S)$ 
                    
                    \Indp Compute Loss, $\mathcal{L}_{\textsf{k}} = \mathcal{L}(\vec{Y}_{\textsf{k}},\hat{Y}_{\textsf{k}})$, and gradient $\frac{\delta{\mathcal{L}_{\textsf{k}}}}{\delta{\hat{Y}_{\textsf{k}}}}$
                    
                    Compute $\frac{\delta{\hat{Y}_{\textsf{k}}}}{\delta{\vec{W}_{n,\textsf{k}}^S}} = \frac{\delta{f_s(\vec{E}_\textsf{k},\vec{W}^S_{n,\textsf{k}})}}{\delta{\vec{W}^S_{n,\textsf{k}}}}$ and $\frac{\delta{\hat{Y}_{\textsf{k}}}}{\delta{\vec{E}_{\textsf{k}}}} = \frac{\delta{f_s(\vec{E}_{\textsf{k}},\vec{W}^S_{n,\textsf{k}})}}{\delta{\vec{E}_{\textsf{k}}}}$

                    Compute gradient $\frac{\delta{\mathcal{L}_{\textsf{k}}}}{\delta{\vec{W}^S_{n,\textsf{k}}}} = \frac{\delta{\mathcal{L}_{\textsf{k}}}}{\delta{\hat{Y}_{\textsf{k}}}} \cdot \frac{\delta{\hat{Y}_{\textsf{k}}}}{\delta{\vec{W}^S_{n,\textsf{k}}}}$ \tcp{Chain rule} 

                    Compute gradient $\frac{\delta{\mathcal{L}_{\textsf{k}}}}{\delta{\vec{E}_{\textsf{k}}}} = \frac{\delta{\mathcal{L}_{\textsf{k}}}}{\delta{\hat{Y}_{\textsf{k}}}} \cdot \frac{\delta{\hat{Y}_{\textsf{k}}}}{\delta{\vec{E}_{\textsf{k}}}}$ \tcp{Chain rule} 
                    
                    Update $\vec{W}^S$ as: $\vec{W}^S_{n,\textsf{k}^\prime} \leftarrow \vec{W}^S_{n,\textsf{k}} - \eta \frac{\delta{\mathcal{L}_{\textsf{k}}}}{\delta{\vec{W}^S_{n, \textsf{k}}}}$

                    \Indm\textit{Quantum Layer:}

                    \Indp Compute $\frac{\delta{\vec{E}_\textsf{k}}}{\delta{\vec{\Theta}_{n, \textsf{k}}}} = \frac{\delta{f_q(\vec{Z}_{\textsf{k}},\vec{\Theta}_{n,\textsf{k}})}}{\delta{\vec{\Theta}_{n,\textsf{k}}}}$ and $\frac{\delta{\vec{E}_{\textsf{k}}}}{\delta{\vec{Z}_{\textsf{k}}}} = \frac{\delta{f_q(\vec{Z}_{\textsf{k}},\vec{\Theta}_{n,\textsf{k}})}}{\delta{\vec{Z}_{\textsf{k}}}}$ \tcp{Parameter-shift rule Eq. \ref{eq:paramshift}, \ref{eq:paramshiftInput}}
                
                    Compute gradient $\frac{\delta{\mathcal{L}_{\textsf{k}}}}{\delta{\vec{\Theta}_{n, \textsf{k}}}}=\frac{\delta{\mathcal{L}_{\textsf{k}}}}{\delta{\vec{E}_{\textsf{k}}}} \cdot \frac{\delta{\vec{E}_{\textsf{k}}}}{\delta{\vec{\Theta}_{n, \textsf{k}}}}$ \tcp{Chain rule} 

                    Compute gradient $\frac{\delta{\mathcal{L_{\textsf{k}}}}}{\delta{\vec{Z}_{\textsf{k}}}} = \frac{\delta{\mathcal{L}_{\textsf{k}}}}{\delta{\vec{E}_{\textsf{k}}}} \cdot \frac{\delta{\vec{E}_{\textsf{k}}}}{\delta{\vec{Z}_{\textsf{k}}}}$ \tcp{Chain rule} 

                    Update $\vec{\Theta}$ as: $\vec{\Theta}_{n, \textsf{k}^\prime} \leftarrow \vec{\Theta}_{n, \textsf{k}} - \eta \frac{\delta{\mathcal{L}_{\textsf{k}}}}{\delta{\vec{\Theta}_{n, \textsf{k}}}}$

                    Send $\frac{\delta{\mathcal{L}_{\textsf{k}}}}{\delta{\vec{Z}_{\textsf{k}}}}$ to client $\textsf{k}$ to continue backward pass
                    
                \Indm \Indm \dotfill{}
                
                \textbf{Client $\textsf{k}$ side:} \tcp{Backward Pass}
    
                    \Indp Compute  $\frac{\delta{\vec{Z}_{\textsf{k}}}}{\delta{\vec{W}^C_{n,\textsf{k}}}} = \frac{\delta{f_c(\vec{X}_{\textsf{k}}, \vec{W}^C_{n, \textsf{k}})}}{\delta{\vec{W}^C_{n,\textsf{k}}}}$ 
    
                    Compute gradient $\frac{\delta{\mathcal{L}_{\textsf{k}}}}{\delta{\vec{W}^C_{n,\textsf{k}}}} = \frac{\delta{\mathcal{L}_{\textsf{k}}}}{\delta{\vec{Z}_{\textsf{k}}}} \cdot \frac{\delta{\vec{Z}_{\textsf{k}}}}{\delta{\vec{W}^C_{n,\textsf{k}}}}$ \tcp{Chain rule} 
    
                    Update client $\textsf{k}$ model weights $\vec{W}^C$ as: $\vec{W}^C_{n, \textsf{k}^\prime} \leftarrow \vec{W}^C_{n, \textsf{k}} - \eta \frac{\delta{\mathcal{L}_{\textsf{k}}}}{\delta{\vec{W}_{n, \textsf{k}}^C}}$

            \Indm\textbf{End of client $\textsf{k}$'s local epoch}

            Send $\vec{W}^{C'}=\vec{W}^C_{n, \textsf{k}^\prime}$ to client $\textsf{k}^\prime$, and start local epoch for client $\textsf{k}^\prime$
            }            
        
        \textbf{End of global epoch $n$}
        
        Start global epoch $n+1$ with client $\textsf{k}$ with updated weights, $\vec{W}^C_{n+1,\textsf{k}^\prime}$, $\vec{W}^S_{n+1,\textsf{k}^\prime}$, and $\vec{\Theta}_{n+1,\textsf{k}^\prime}$
        }
    \textbf{END OF TRAINING}

    \Return $\vec{\tilde{W}}{=}(\vec{\tilde{W}}^C,\vec{\tilde{W}}^S)$, $\vec{\tilde{\Theta}}$
\end{algorithm}
\section{\label{sec:hqsl_exp}Experiments and Results}

In this section, we empirically assess the feasibility and scalability of HQSL by comparing its performance in classification tasks to that of their corresponding SL models. We first outline our experiments and then present our results and discuss our findings.
All our programs were written using Python 3.11.3 and PyTorch 2.1.0 libraries. We simulate the quantum part of our HQSL models using the Pennylane library with its PyTorch backend. The experiments were conducted on an NVIDIA GeForce RTX 2080 Ti GPU machine system. 

We use five publicly available datasets in this work to test and validate our HQSL architectures. We summarize these datasets in Table \ref{tab:datasets}. Detailed descriptions, including dataset processing methods, can be found in Appendix \ref{appendix:datasets}. 
\begin{table}[htbp!]
    \centering
    \caption{Datasets used for our experiments}
    \label{tab:datasets}
    \begin{tabular}{ccccc}
    \toprule
         \multirow{2}{*}{Dataset}  & Training & Testing  & $\#$ features/ & \multirow{2}{*}{$\#$ classes} \\
         & samples & samples & Image size &\\
         \hline
         Botnet DGA \citep{IEEEPort}  & 800 & 200 & 7 & 2\\
         Breast Cancer \citep{misc_breast_cancer_14} & 455 & 114&7 & 2\\
         MNIST \citep{deng2012mnist}  & 4800 & 1200 & 28 x 28&10\\
         Fashion-MNIST \citep{xiao2017fashionmnist}  & 4800 & 1200 & 28 x 28 & 10\\
         Speech Commands \citep{warden2018speech} & 6388 & 1597 & 28 x 28 & 2\\
         \botrule
    \end{tabular}
\end{table}
Setting a fixed seed, we split our datasets into five non-overlapping folds, in preparation for five-fold cross-validation for our experiments. Specifically, we use four of the folds for collectively training our model, and the remaining fold to test it. For the multi-class image datasets (MNIST and Fashion-MNIST), we split the datasets in a stratified manner to ensure an even and balanced distribution of classes. 
\subsection{\label{subsec: exp_1client_tests} Hybrid Quantum Split Learning versus Classical Split Learning: Single Client Experiments}
In this section, we describe our experiments to determine  HQSL's feasibility and performance by considering a single client-single server case. We then discuss the results of our experiment.

First, we constructed the quantum circuit shown in Fig. \ref{fig:circuit} using PennyLane's \code{default-qubit} device and converted it to a quantum layer. 
To build HQSL variants 1 and 2 as introduced in Section \ref{subsubsec:HQSL_modelling}, we first constructed their respective client-side and server-side model portions separately, as illustrated in Fig. \ref{fig:HQSLmodels}, ensuring that each variant adhered to the split learning framework. We emphasize again here that the quantum layer was incorporated as the initial layer of the server-side model portion of each HQSL variant. This separation allows for the collaborative training of the client-side and server-side model portions in HQSL while maintaining the distinct roles of the client and the server as per the split learning protocol. 
 
We built HQSL variant 1 for training on Botnet DGA and Breast Cancer datasets and HQSL variant 2 on MNIST, Fashion-MNIST, and Speech Command spectrogram datasets. In HQSL variant 2, we set the number of output nodes to $n=10$ (10 classes) for the MNIST and Fashion-MNIST datasets and $n=2$ (2 classes) for the Speech Commands spectrograms dataset. We paired each of these 5 experiments with their classical counterparts to benchmark the performance of our HQSL model variants.

For our hyperparameters, we utilized the \code{Adam} optimizer with a learning rate of $10^{-3}$ across all experiments, except for the Speech Commands dataset, where a learning rate of $10^{-4}$ provided better convergence. We used the binary cross entropy loss function for classifying the 2 multi-variate datasets and the cross entropy loss function for the 3 image classification tasks. We trained our HQSL model variants as per Algorithm \ref{alg:hybsplit} for 100 epochs for Botnet DGA and Breast Cancer datasets and 50 epochs for MNIST, FMNIST, and Speech Commands spectrograms datasets.
We report the mean and standard deviation of the accuracies and F1-scores over five folds of the datasets in Fig. \ref{fig:feasibility}. 
Across all pairs of experiments, we set the random seed to be a constant and fixed the training and testing datasets for each of the 5 folds. These experiments were also performed on the centralized (unsplit) versions of each split learning model. We highlight here that the training process of the centralized version of HQSL is identical to that of a single-node HQNN. We next compare the classification performance results of single-client HQSL against their classical SL equivalent, as discussed in the next section. 

\noindent
\textbf{Hardware and Noisy Simulator Experiments.}
To evaluate the feasibility of deploying HQSL on real or noisy quantum hardware, we tested the model on both real IBMQ devices and simulated noisy environments using \texttt{qiskit\_aer}~\citep{qiskit2024} with increasing single and two-qubit depolarizing noise levels ($p=0.05$ to $0.09$). We reference calibrated error rates from the 127-qubit \texttt{ibm\_brisbane} backend, where median single-qubit gate errors (SX) are typically below $3e^{-4}$, and two-qubit gate (ECR) errors have a median of $7e^{-3}$, which are substantially below the noise levels used in our noisy simulation experiments. These simulations allow us to test the limits of HQSL’s robustness under exaggerated noise conditions that exceed those observed on current hardware. We benchmarked against Pennylane's noise-free \texttt{default.qubit} device. Due to the high execution time associated with accessing real quantum hardware, we limited our evaluation to a single fold of the Botnet DGA dataset. Specifically, we trained single-client HQSL using the training set on the \texttt{default.qubit} simulator and evaluated its performance on the various backends using the corresponding testing set. These results are presented in Fig.~\ref{fig:hardware}.

\subsubsection*{\label{1_client_result} Results and Discussions}
\begin{figure}[htbp!]
 \includegraphics[width=\textwidth]{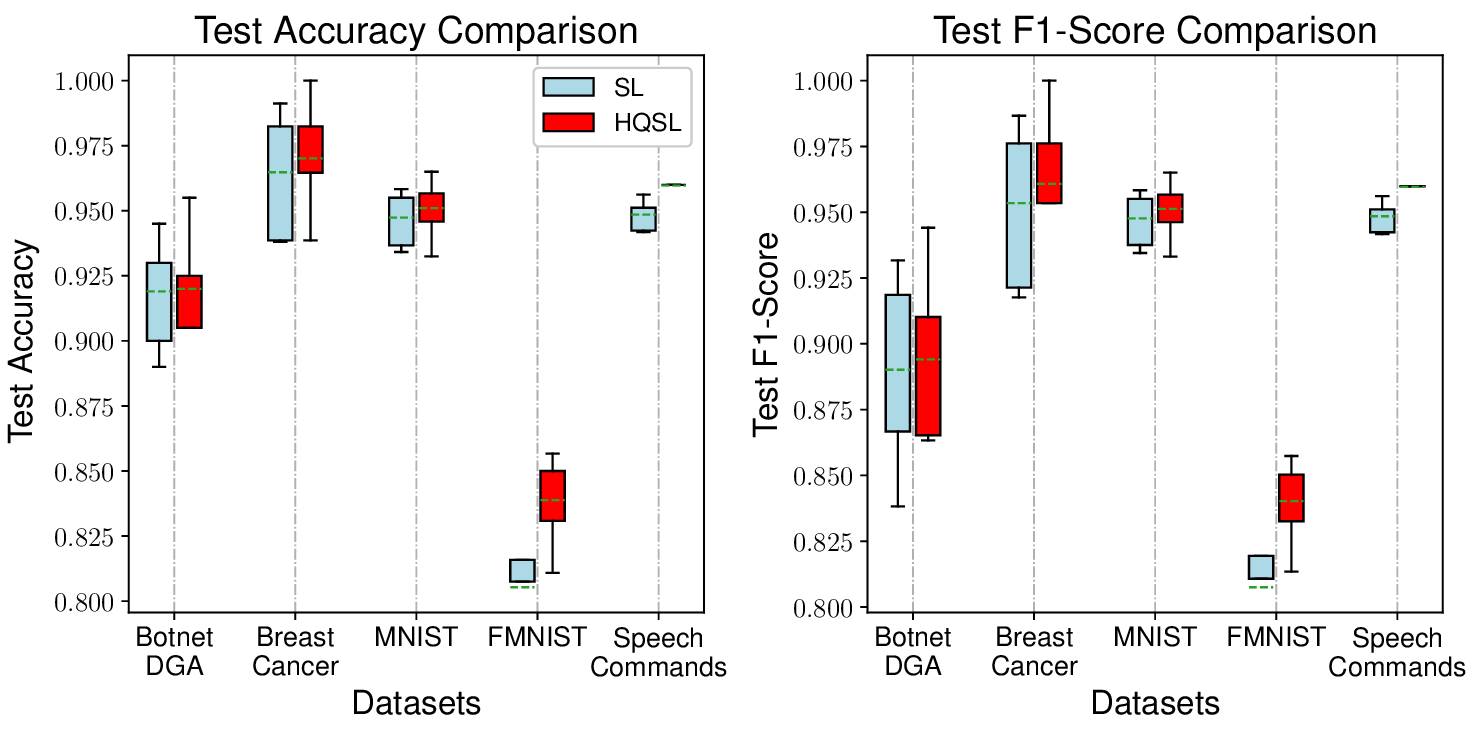}
\caption{Performance comparison in terms of test accuracy and F1-score of SL (\textcolor{CornflowerBlue}{blue}) versus HQSL (\textcolor{red}{red}) with 5-fold cross-validation. The dashed green lines spanning the width of each box represent the mean test results. In every case, HQSL outperforms or is nearly on par with SL. On the FMNIST and Speech Commands datasets, a mean improvement of approximately 3\% and 1.5\%, respectively, due to HQSL is observed}
\label{fig:feasibility}
\end{figure}

We present the results of our tests on HQSL and SL for a single client in Fig. \ref{fig:feasibility}. We highlight that the test results for the split and centralized HQSL versions are identical, confirming that the splitting process does not degrade model accuracy and F1-score. This demonstrates that HQSL successfully maintains the benefits of split learning (collaborative training of a resource-constrained client with a quantum-enhanced server) while achieving the same performance as its centralized single-node HQNN counterpart.
By applying the same approach to their equivalent SL model, we conduct a comparative analysis of each model's performance in terms of accuracy and F1-score. Fig. \ref{fig:feasibility} illustrates the 5-fold test accuracies and F1-scores obtained across all datasets in box plot format. The mean values are represented by the dashed green line within each box, while the heights of the boxes indicate the standard deviations in the testing results.

Although marginal, improvements in both mean accuracies and F1-scores of HQSL were obtained on the Botnet DGA (mean accuracy improvement: $\uparrow0.1\%$, mean F1-score improvement: $\uparrow0.4\%$), Breast Cancer ($\uparrow0.6\%$, $\uparrow0.7\%$) and MNIST ($\uparrow0.4\%$, $\uparrow0.4\%$) datasets.  
Significant outperformance in terms of mean accuracy and F1-score were obtained on the Fashion-MNIST ($\uparrow3\%$, $\uparrow4\%$) and Speech Commands ($\uparrow1.5\%$, $\uparrow1.5\%$) datasets. The visually narrow box plot for the Speech Commands dataset in the HQSL case is a result of the low variance across the five cross-validation folds. This indicates the higher stability of our HQSL model's performance on this dataset compared to the other datasets and to SL's performance of the Speech Commands dataset. For completeness, we provide the performance results for the Speech Commands dataset in Appendix \ref{appendix:speech}.

For all of the five datasets used in this study, HQSL proved feasible as it at least matched the testing performance of SL in terms of accuracies and F1-scores. As demonstrated by the results on the  Fashion-MNIST and Speech Commands spectrograms datasets, HQSL has the potential to achieve higher accuracy and F1-score compared to its classical counterpart.
We highlight that with just the introduction of 1 quantum layer consisting of a small 2-qubit quantum circuit, HQSL can slightly outperform classical split learning, showcasing the potential power of quantum computation beyond the NISQ era.

It is noteworthy that simulating a quantum computer using a classical computer is computationally intensive, despite having only 1 quantum layer consisting of a simple quantum circuit on the server side of HQSL. The training time for HQSL was significantly longer than that of SL, depending on the size of the dataset. For instance, 5-fold training for 100 epochs, each with 569 data points from the Breast Cancer dataset split in the train:test ratio of 4:1 took approximately 1 hour to run on HQSL. On the other hand, 5-fold training with 6000 image-label pairs from the MNIST dataset took around 9 hours to complete 50 training epochs of HQSL. In contrast, their classical analogues only took a few minutes to train which is significantly faster than HQSL. 

\begin{figure}[htbp!]
 \includegraphics[width=\textwidth]{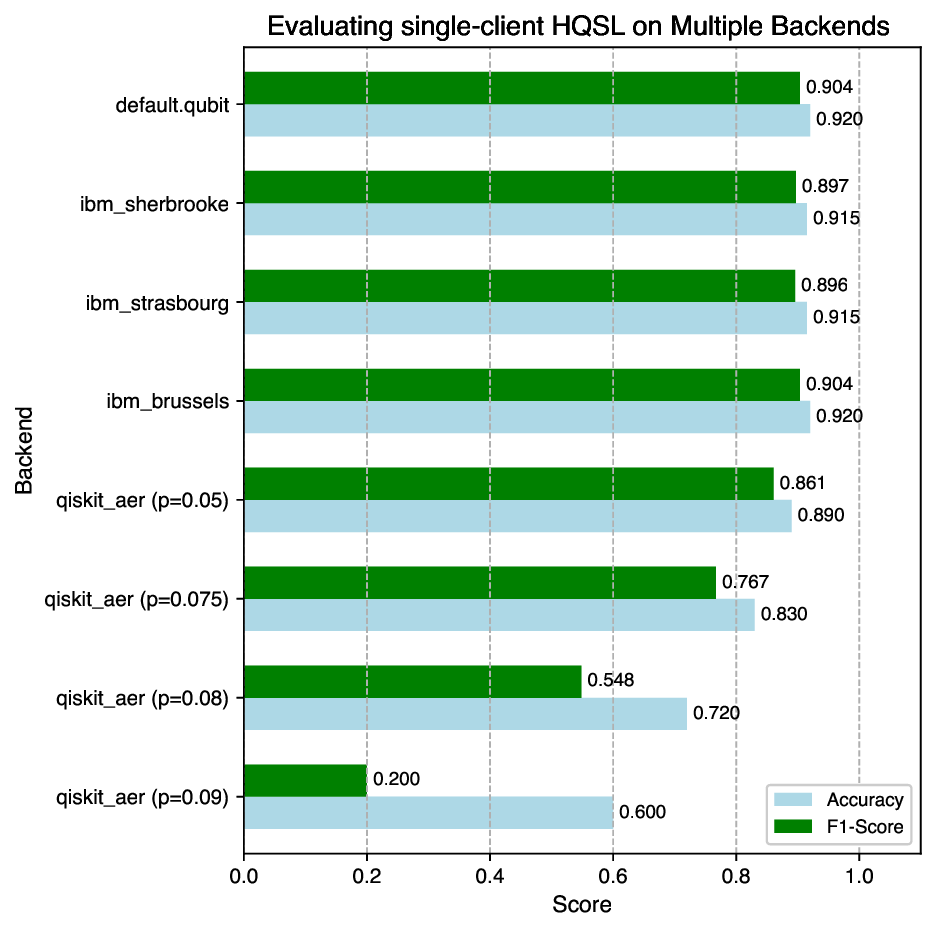}
\caption{Deployment accuracy and F1-score of HQSL evaluated on the Botnet DGA dataset across multiple backends. \texttt{default.qubit} represents HQSL's noise-free simulation results, which match the performance obtained on IBMQ hardware (\texttt{ibm\_sherbrooke}, \texttt{ibm\_strasbourg}, \texttt{ibm\_brussels}). Noisy simulators (\texttt{qiskit\_aer}) with increasing depolarizing noise $p$ demonstrate a drop in performance as noise levels increase. This shows HQSL's noise-resilience during deployment on currently available hardware}
\label{fig:hardware}
\end{figure}

\noindent
\textbf{Evaluation Results on Real Quantum Devices and Noisy Simulators.}
To assess the practical viability of HQSL under real-world noise, we evaluated the model on both real IBMQ backends and noisy simulators. As shown in Fig.~\ref{fig:hardware}, HQSL achieves consistently high accuracy and F1-scores on \texttt{ibm\_sherbrooke}, \texttt{ibm\_strasbourg}, and \texttt{ibm\_brussels}, similar to the noise-free \texttt{default.qubit} simulator, indicating that currently available noisy quantum hardware is sufficient to support HQSL deployment with our proposed qubit-efficient data-loading quantum circuit in its quantum layer. Interestingly, significant performance drops occurred only when using unrealistically high depolarizing noise models in simulation (e.g., $p = 0.08$ or $p = 0.09$), suggesting that these levels represent overly pessimistic device conditions. These results further support the feasibility of deploying HQSL on real hardware.

\noindent
\textit{In the single client case, HQSL with a quantum layer consisting of only a small quantum circuit is feasible and offers tangible improvements in classification accuracy and F1-score compared to SL. Moreover, evaluations on real quantum hardware and noisy simulators highlight HQSL’s practical feasibility and robustness for deployment on currently available quantum devices.}
\subsection{\label{subsec:exp_scalability}Hybrid Quantum Split Learning versus Classical Split Learning: Multiple-client Experiments}
In this section, we experiment with HQSL with an increasing number of clients to determine its potential for accommodating multiple clients. We then discuss our empirical findings comparing HQSL and SL in a multiple-client setup.

First, to assess the impact of the number of clients $\textsf{K}$ on model accuracy and F1-score, we divide the training set into $\textsf{K}$ independent and identically distributed (IID) subsets for $\textsf{K} \in\{2, 3, 4, 5, 10, 20, 50, 100\}$. Each subset represents an IID distribution of the dataset assigned to each client, ensuring a balanced portion for each client. 

The training process follows Algorithm \ref{alg:hybsplitN_Clients}. The hyperparameters used are the same as in the case of a single client in Section \ref{subsec: exp_1client_tests}. At the end of each global epoch, we evaluate the model on the testing set. We repeat this training process for 100 global epochs on the multivariate datasets (Botnet DGA and Breast Cancer) and 50 global epochs on the image datasets (MNIST, FMNIST, and Speech Commands spectrograms). The test results are then compared to those of their classical equivalent. The reported results include the test accuracy and F1-score at the end of training comparing HQSL against SL.
In this set of experiments, these tests are conducted for only one fold, and the test accuracies and F1-scores are reported to evaluate the scalability of HQSL compared to SL.

\subsubsection*{Results and Discussions}
\begin{figure}[!htbp]
\includegraphics[width=\textwidth]{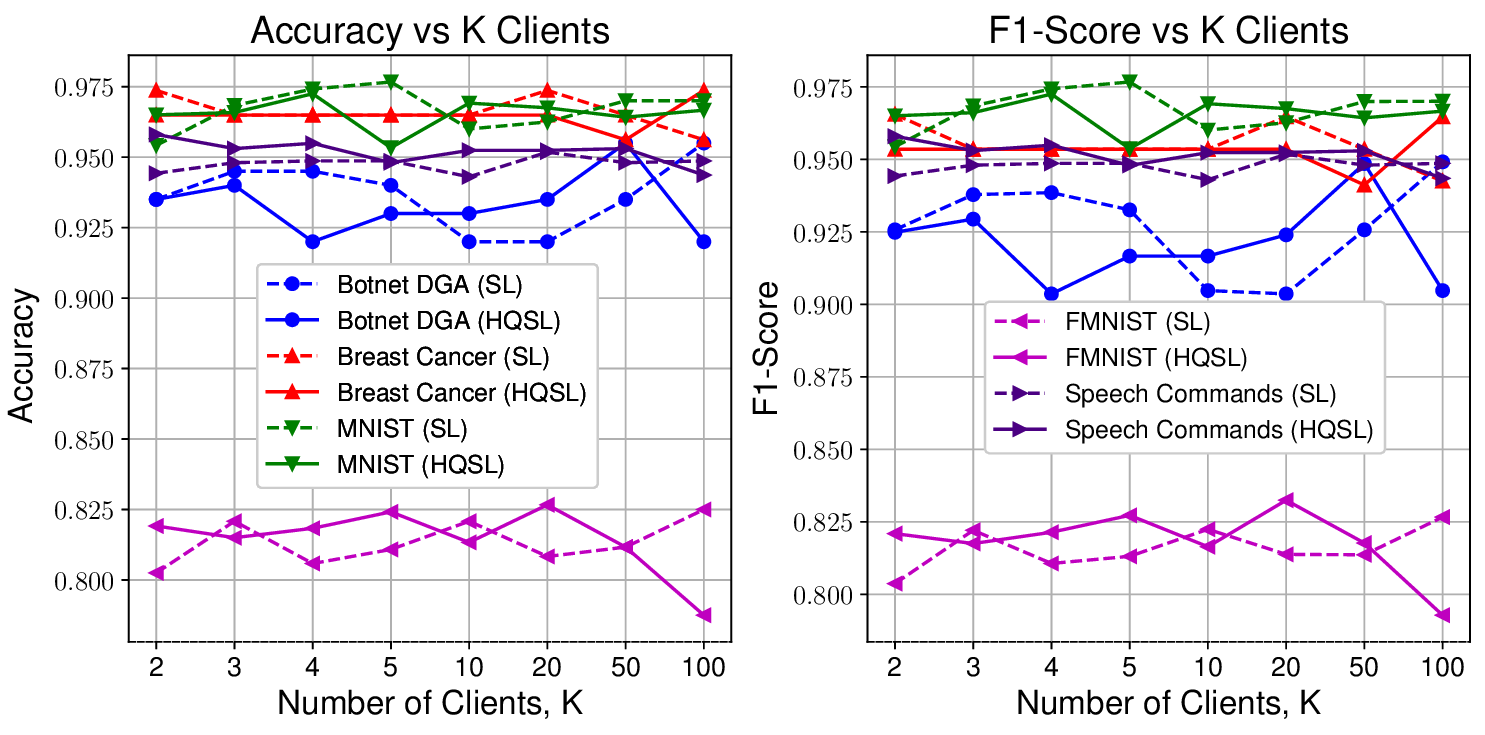}
\caption{Test accuracy and F1-score performances of HQSL compared with SL on all 5 datasets for up to $\textsf{K}=100$ clients. HQSL maintains a high performance like SL while accommodating multiple clients}
\label{fig:scalability}
\end{figure}

Fig. \ref{fig:scalability} depicts the accuracy and F1-score of HQSL as the number of clients $\textsf{K}$ increases compared to SL. 
For both HQSL and SL, the accuracy and F1-score do not show a significant drop in performance across all datasets as we increase the number of clients, $\textsf{K}$. The result suggests that HQSL can match, and in some cases potentially surpass, the performance of traditional SL as the client count increases. This finding underscores the viability of HQSL.

While our results are promising, we acknowledge the presence of some fluctuations in the testing outcomes. These variations can be attributed to our current experimental design, which utilized a single fold of the dataset. 
Although we cannot assert HQSL's superiority over SL in multi-client scenarios based on the current result, the evidence strongly supports HQSL's feasibility and scalability. The comparable performance between HQSL and SL across increasing client numbers is a significant finding, opening up new avenues for quantum-enhanced distributed learning in real-world applications.

We used an IID dataset distribution among our $\textsf{K}$ clients to ensure an even and balanced distribution of the dataset for each client. Moreover, in this study, we only considered the round-robin communication protocol for training each client with a centralized server. However, there are more sophisticated methods available in academic literature, such as asynchronous training.
In addition, increasing the number of clients increases the training latency for a particular client in the network, which is further exacerbated by the round-robin communication protocol we use.
While this study of HQSL's scalability with multiple clients does not address cases involving unbalanced dataset distribution across clients, more advanced communication protocols among clients, or the reduction of the training latency of each client, we defer these for future research.

\textit{By leveraging the scalability property of split learning, HQSL can accommodate multiple clients without dropping performance. The performance trend is similar to that of the SL counterpart.}

\section{\label{sec:reconstruction}Enhancing the Resistance of HQSL against Reconstruction Attacks}
In this section, we address the second research question \textit{RQ 2} stated as follows: \textit{How do we strengthen such hybrid quantum split learning schemes against data privacy leakage and reconstruction attacks?}
Specifically, we investigate the susceptibility of HQSL to reconstruction attacks using smashed data on the server side. First, we introduce the threat model, outlining the potential risks and vulnerabilities of HQSL to reconstruction attacks. Then, we present the reconstruction attack models we consider in this work to recreate a client's raw input data. These models are designed to exploit the smashed data on the server side.

Considering the rotational properties of the encoding gates in our quantum layer, we propose a noise-based defense mechanism in the form of a Laplacian noise layer. The split structure of HQSL conveniently allows us to insert a noise layer at the client-server boundary, directly on the transmitted smashed data. Specifically, this layer adds controlled randomness to the smashed data before they are transmitted to the server-side model, making it more difficult for the reconstruction attack models to recreate private input raw data. Next, we conduct experiments under different noise parameter settings to extract the best-performing noise layer whereby HQSL has a clear advantage over SL in two key aspects: (i) impairing the performance of the reconstruction attack models and (ii) maintaining a high classification performance despite the presence of the noise defense layer. In our experiments, we investigate reconstruction attacks considering the image datasets introduced in Section \ref{sec:hqsl_exp}, specifically, MNIST, Fashion-MNIST, and the spectrograms from the Speech Commands datasets.
\subsection{\label{subsec: rec_methodology}Reconstruction Attack from Smashed Data in HQSL}
In this section, we present our threat model that describes how an honest but curious server adversary can recreate the private input raw data from the smashed data transmitted by the client-side model. The adversary uses a reconstruction attack model to recreate the client's input data at inference time. Hence, we subsequently describe three such reconstruction attack models that can be employed by the adversary. We then describe how we design our proposed defense mechanism to mitigate the risk of reconstruction attacks in HQSL. This countermeasure entails the introduction of a Laplacian noise layer at the end of the client-side model, designed and configured considering the rotational properties of encoding gates in the server-side quantum layer.
\subsubsection{\label{subsubsec: rec_threat_model}Threat Model}
In this work, we operate under the premise that we have a server that collaborates with multiple clients or data owners according to the SL setup.  
The clients do not share their private dataset, $\mathcal{D}_\textsf{priv}$, with the server or other clients in the network.  We assume that the server is honest but curious (semi-honest), i.e., the server does not deviate from the specified protocol instructions but attempts to infer information about the client's data. 
We assume that the server has access to a publicly available auxiliary dataset, $\mathcal{D}_\textsf{aux}$, that has a similar distribution as $\mathcal{D}_\textsf{priv}$, but $\mathcal{D}_\textsf{priv}$ and $\mathcal{D}_\textsf{aux}$ are non-overlapping disjoint datasets. The access to such a dataset is a common assumption made by previous works \citep{dougherty_stealthy_2023, pasquini_unleashing_2021, shokri_membership_2017}. 
As explained in \citep{pasquini_unleashing_2021}, the adversary can have access to a `shadow' model, $\textsf{f}_\textsf{shadow}$,  to generate outputs, $\textsf{Z}_\textsf{aux}$, in the same feature space as the smashed data, $\textsf{Z}_\textsf{priv}$, received from a client. The shadow model is trained on the auxiliary dataset, $\mathcal{D}_\textsf{aux}$. The server then uses the pair 
$(\textsf{Z}_\textsf{aux}=\textsf{f}_\textsf{shadow}(\textsf{X}_\textsf{aux}), \textsf{X}_\textsf{aux})$, where $\textsf{X}_\textsf{aux} \subset \mathcal{D}_\textsf{aux}$ to train a reconstruction attack model, $f_\textsf{rec}$, that recreates $\textsf{X}_\textsf{aux}$. 

With the trained reconstruction attack model, $\textsf{f}_\textsf{rec}$, and the smashed data, $\textsf{Z}_\textsf{priv}$, generated from the private dataset, $\mathcal{D}_\textsf{priv}$, the honest-but-curious server reconstructs the private input of the clients. \textit{In this work, we assume that reconstruction attacks happen at deployment/inference time. This work does not address split learning's susceptibility to data privacy leakage and reconstruction attacks during the training phase.} 
\subsubsection{\label{subsec:attack_model}Reconstruction Attack Mechanism}
\begin{figure*}
    \centering
    \includegraphics[width = \textwidth]{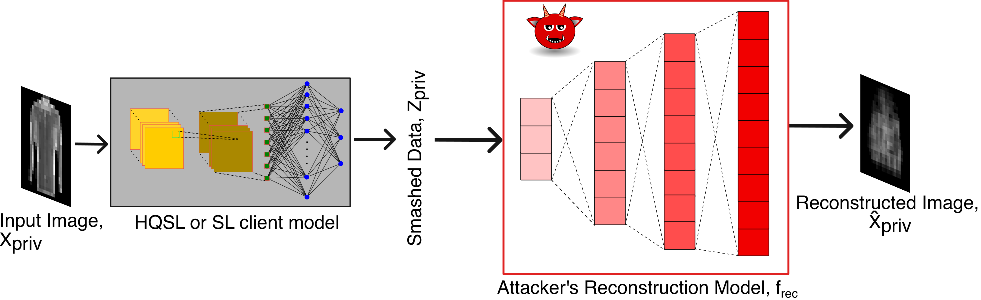}
    \caption{Setup for reconstruction attacks using $\textsf{Z}_\textsf{priv}$ obtained from either HQSL or SL's client model to reconstruct $\textsf{X}_\textsf{priv}$. $\hat{\textsf{X}}_\textsf{priv}$ is the reconstructed version of $\textsf{X}_\textsf{priv}$}    \label{fig:attack_model}
\end{figure*}

We consider three distinct reconstruction model architectures to reconstruct the private input data, $\textsf{X}_\textsf{priv} \subset \mathcal{D}_\textsf{priv}$, during a reconstruction attack using the smashed data, $\textsf{Z}_\textsf{priv}$. We train the reconstruction models on the auxiliary dataset, $\mathcal{D}_\textsf{aux}$. During training, the inputs to these models are the smashed data, $\textsf{Z}_\textsf{aux}$, and the outputs are the reconstructed images, $\hat{\textsf{X}}_\textsf{aux}$. These models are obtained from the literature, and we will assess the performance of our proposed defense mechanism (see Section \ref{subsubsec:defence_model}) against these trained attack models on both HQSL and SL.
Since the dimension of the smashed data, $dim(\textsf{Z}_\textsf{priv}) = 3$ and the dimension of the client's input data, $dim(\textsf{X}_\textsf{priv}) = 28$ x $28$ for the image datasets (MNIST, FMNIST, and Speech Commands spectrograms), we adjust the input and output layer dimensions of the reconstruction models. We describe the three attack models we investigate in this work as follows:
\begin{enumerate}
    \item \textbf{Reconstruction Model 1}: We use the reconstruction attack model architecture proposed by Vepakomma et al. \citep{vepakomma_nopeek_2020}, which performs feature inversion to reconstruct images from low dimensional intermediate features. The model is a neural network with transpose convolutions. It consists of skip connections such as those used in \code{ResNet} models, with \code{ReLU} activation and batch normalization. 
    \item \textbf{Reconstruction Model 2}: Li F. et al. \citep{li2018discriminatively} proposed a fully convolutional autoencoder for image feature extraction. We adopt the decoder part of the autoencoder to reconstruct the input images. Upsampling layers are used to recover the feature maps along with deconvolutional layers with \code{ReLU} activation function. Batch normalization is used after each deconvolutional layer except for the last layer. The Euclidean/\code{MSELoss} loss function is used to compute the difference between the original and reconstructed images. We use a learning rate of 0.001 and \code{Adam} optimizer to train Reconstruction Model 2.
    \item \textbf{Reconstruction Model 3}: In \citep{balle2022reconstructing}, a fully connected model consisting of 2 hidden layers with a width of 1000 and the \code{ReLU} activation function is used as the decoder part of the autoencoder. The loss function used here is the mean absolute error (MAE/\code{L1Loss}) + mean square error (\code{MSELoss}). A learning rate of 0.001 with the \code{RMSprop} optimizer is used to train Reconstruction Model 3.
\end{enumerate}
After training the reconstruction models, they are deployed to recreate a client's private input. In Fig. \ref{fig:attack_model}, we illustrate the setup of a reconstruction attack, on the private data from a client, $\textsf{X}_\textsf{priv}$, using their smashed data, $\textsf{Z}_\textsf{priv}$, to successfully recreate their private input data, represented by $\hat{\textsf{X}}_\textsf{priv}$. 

The smashed data, $\textsf{Z}_\textsf{priv}$, transmitted to the server model consists of latent information about the client's private input data, $\textsf{X}_\textsf{priv}$, which can be reconstructed by the setup shown in Fig. \ref{fig:attack_model}. In Section \ref{subsubsec:metrics}, the baseline results for the image comparison metrics demonstrate high similarities between the original ($\textsf{X}_\textsf{priv}$) and reconstructed images ($\hat{\textsf{X}}_\textsf{priv}$). Hence, to maintain the privacy of the client's input data, the smashed data communicated to the server model must be desensitized to reduce the possibility of successful reconstruction attacks. Thus, we need a defense mechanism to mitigate the risks of reconstruction of private input raw images from smashed data in HQSL. 

\subsubsection{\label{subsubsec:defence_model}Encoding Gate-based Noise Layer Defense Mechanism}
To defend HQSL against the risk of reconstruction attacks, we propose a Laplacian noise layer defense mechanism at the end of the client-side model. Previous works have considered the use of a noise layer defense mechanism to introduce perturbation to the smashed data and hence, increase the difficulty of reconstruction by the proposed reconstruction models. However, such noise defense methods come with a privacy-utility trade-off \citep{titcombe_practical_2021, na2023systematic, mireshghallah_shredder_2020}. In this section, we study the rotational properties of quantum encoding gates to design the noise layer to
(i) effectively hinder the reconstruction attack, and (ii) maintain a high classification performance (accuracy and F1-score) by the server during inference. This, hence, addresses the privacy-utility trade-off associated with the noise layer defense mechanism.
We empirically compare these two aspects for the hybrid (HQSL) and classical (SL) cases in Sections \ref{subsec: rec_exp} and \ref{subsec: ResultsReconstruction}. 

Laplacian noise is favored over other types of noise, such as Gaussian noise, for privacy protection or data obfuscation. 
This is because noise sampled from a Laplacian distribution, $\mathrm{Laplace}(\mu, b)$, can be systematically added to the smashed data, $\textsf{Z}_\textsf{priv}$, to provide controlled randomness, while enabling the explicit quantification of the level of perturbation introduced. This is important for maintaining privacy while retaining data utility. Laplacian noise is also commonly utilized in the context of differential privacy \citep{shokri_privacy-preserving_2015, sarathy2011evaluating}. 

\begin{figure}[htbp!]
    \centering
    \begin{subfigure}{0.45\textwidth}
    \centering
        \includegraphics[width=\textwidth]{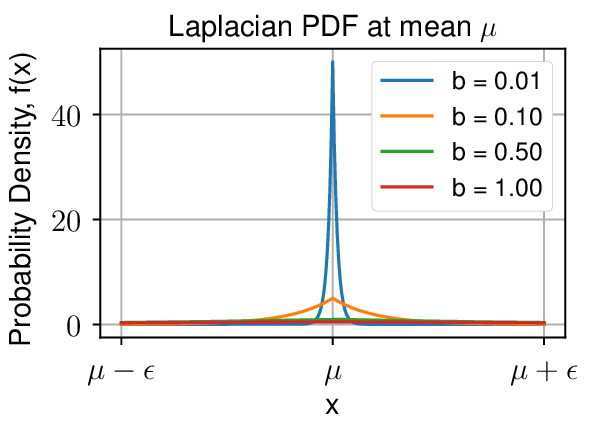}
        \caption{Probability density function (PDF) of Laplacian noise centered at mean $\mu$ with varying scale parameter, $b$}
        \label{fig:Laplacian_Dist}
    \end{subfigure}
    \hskip4pt
    \begin{subfigure}{0.5\textwidth}
    \centering  
        \includegraphics[width=\textwidth]{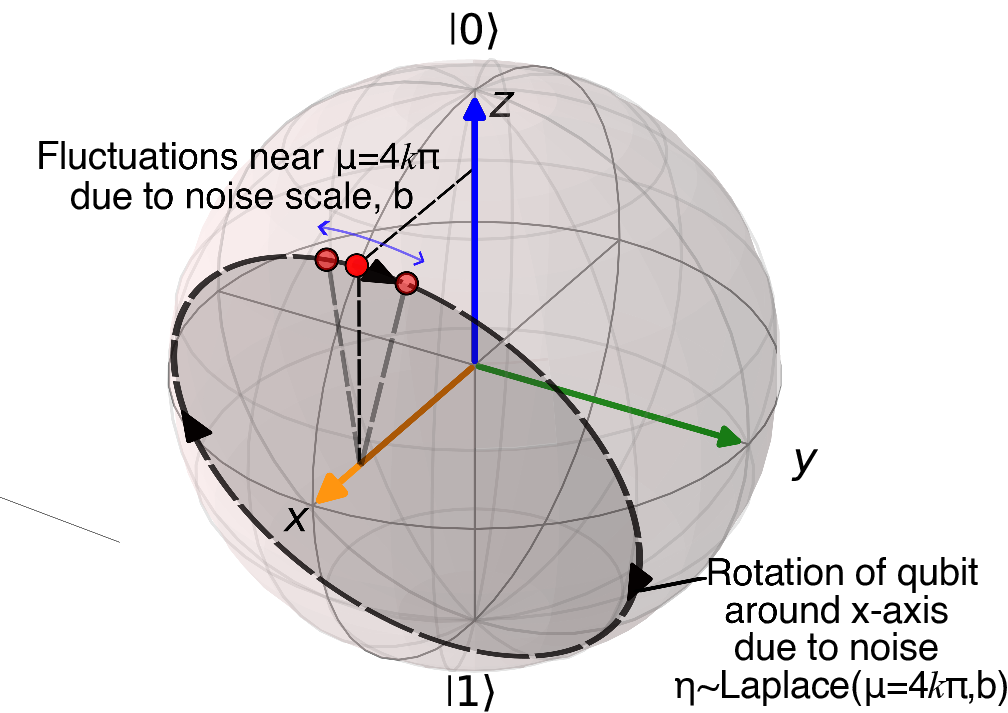}
        \caption{Rotation of qubit (\textcolor{red}{red dot}) around the x-axis of the Bloch sphere after adding Laplacian noise, $\eta$, with mean $\mu=4k\pi$ and small scale, $b$}
        \label{fig:Bloch_sphere}
    \end{subfigure}
    \caption{(a) As the scale value $b$ decreases, the Laplacian noise becomes more concentrated around the mean, $\mu$, resulting in a higher density of data points near $\mu$. (b) The addition of this noise (mean $\mu=4k\pi$, $k\in\mathbb{Z}$, and small scale $b$) to the smashed data causes only a small fluctuation near the original location (\textcolor{red}{red dot}) of the qubit about the x-axis on the Bloch sphere}
    \label{fig:b_sphere}
\end{figure}

The location (mean) parameter $\mu$ shifts the distribution, while the scale parameter $b$ controls its spread. In our setting, this is critical: the scale parameter determines how tightly noise samples concentrate around the mean. As shown in Fig. \ref{fig:Laplacian_Dist}, smaller $b$ values result in a sharper distribution centered at $\mu$, keeping the noise values sampled close to the desired noise level. This sharpness allows us to maintain high fidelity in the quantum state encoding while still introducing enough variability to hinder adversarial reconstruction, as we show next.

\vspace{0.5em}
\noindent
\textbf{Noise Layer Design Based on Rotational Properties of RX-gates.}

\noindent
Adding noise sampled from a Laplacian distribution, with appropriately tuned mean and scale parameters, has the potential to benefit the quantum circuit on the server side by maintaining high classification performance while impeding reconstruction of raw data via the added jitter.  In Section \ref{subsec: ResultsReconstruction}, we will demonstrate that a purely classical server cannot simultaneously achieve these goals.

We consider the following to develop a better understanding of our selection of Laplacian noise parameters. The Bloch sphere gives a geometrical representation of the pure‐state space of a qubit.  Single‐qubit rotation gates, e.g., the RX-gate, move the state vector around the surface of the Bloch sphere, as shown in Fig. \ref{fig:Bloch_sphere}.  In our proposed HQSL circuit (Fig. \ref{fig:circuit}), each RX‐gate has a $4\pi$ period and encodes a classical feature $z$ by rotating the qubit about the x‐axis by angle $\theta=z$.  Adding noise $\eta$ to the smashed data, hence, $\tilde{\theta} = z + \eta$, corresponds to applying

\begin{equation}
      RX(\tilde\theta)\;=\;RX(z+\eta)\,.
\end{equation}
By the $4\pi$ periodicity of $RX(\theta)$, centering our Laplace noise at any mean $\mu=2k\pi$ ($k\in\mathbb Z$) ensures that as $b\to0$,
\begin{equation}
  RX(z+\eta)\;\approx\;RX(z)
  \quad(\text{up to a global phase}),    
\end{equation}
so that the encoded quantum state is preserved. By computing the fidelity between the clean and noisy encoded quantum states, we formalize this below.

\noindent
\textit{Expected Fidelity Preservation.}

\noindent
Using the matrix representation of our encoding RX-gate, the encoded quantum state, $\lvert\psi(z)\rangle$, can be written as follows:
\begin{equation}
  \lvert\psi(z)\rangle = RX(z)\lvert0\rangle
    = \cos\!\tfrac{z}{2}\lvert0\rangle - i\sin\!\tfrac{z}{2}\lvert1\rangle,
\end{equation}
and with noise $\eta\sim\mathrm{Laplace}(0,b)$ we have:
\begin{equation}
  \lvert\psi(z+\eta)\rangle
    = \cos\!\tfrac{z+\eta}{2}\lvert0\rangle - i\sin\!\tfrac{z+\eta}{2}\lvert1\rangle.
\end{equation}
The state fidelity is
\begin{equation}
  F(\eta) = \bigl|\langle\psi(z)\mid\psi(z+\eta)\rangle\bigr|^2
    = \cos^2\!\tfrac{\eta}{2}
    = \tfrac{1+\cos\eta}{2}.
\end{equation}

Taking expectation over $\eta$ and using the Laplace characteristic function yields
\begin{equation}
  \mathbb{E}_\eta[F]
    = \frac12\Bigl(1 + \mathbb{E}[\cos\eta]\Bigr)
    = \frac12\Bigl(1 + \tfrac{1}{1 + b^2}\Bigr).
\end{equation}

In particular, for $b\ll1$, we have $\mathbb{E}_\eta[F]\approx1$. This implies that the clean and noisy encoded quantum state on the server-side quantum layer remains unchanged as $b\to0$.
Hence, keeping the scale parameter $b$ small limits the noise to a minimal perturbation about the mean, as illustrated by the sharp Laplacian PDF in Fig. \ref{fig:Laplacian_Dist}. This low variability ensures consistent encoding that closely matches the clean quantum state, preserving the hybrid server model’s utility.

\begin{figure*}[htpb!]
    \centering
    \includegraphics[width=\textwidth]{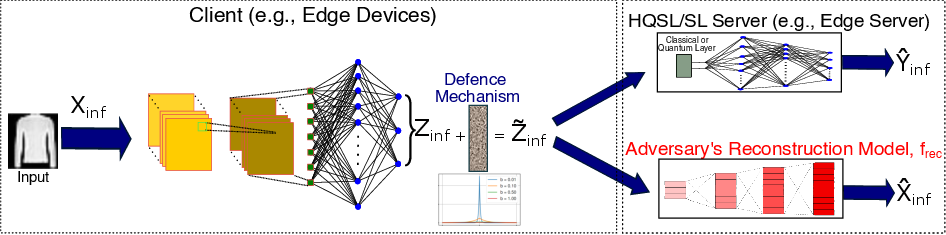}
    \caption{At inference time, $\textsf{X}_\textsf{inf}$ is the input to the client model with the Laplacian noise layer as the defense mechanism. Laplacian noise is added to the smashed data, $\textsf{Z}_\textsf{inf}$, resulting in $\tilde{\textsf{Z}}_\textsf{inf}$. $\tilde{\textsf{Z}}_\textsf{inf}$ is communicated to the HQSL or SL server to carry out the classification task with outputs $\tilde{\textsf{Y}}_\textsf{inf}$ as intended or fed to a reconstruction model to recreate private input data as $\hat{\textsf{X}}_\textsf{inf}$}
    \label{fig:ARM_model}
\end{figure*}

In Fig. \ref{fig:ARM_model}, we illustrate the defense setup at inference time: the client’s smashed data $\textsf{Z}_\textsf{inf}$ is perturbed to $\tilde{\textsf{Z}}_\textsf{inf}$ via the Laplacian noise layer before entering the server’s quantum circuit, thereby obscuring the input against reconstruction attacks while retaining classification accuracy.
We also include Algorithm \ref{alg:hqsl_inference} to outline the deployment of the Laplacian noise-based defense mechanism at inference time as a countermeasure against reconstruction attacks on the server side. The algorithm describes the process of obfuscating the smashed data $\textsf{Z}_\texttt{inf}$ before sending $\Tilde{\textsf{Z}}_\texttt{inf}$ to the HQSL server. Depending on the selected mode, the server either performs classification using the hybrid quantum server model or attempts to reconstruct the original input $\textsf{X}_\texttt{inf}$.

\begin{algorithm}[htbp!]
    \caption{HQSL with Laplacian Noise Defense Deployed at Inference Time}
    \label{alg:hqsl_inference}
    \textbf{Input:} Pre-trained Client Model $f_c$ with weights $\vec{\Tilde{W}}^C$, Server Model with classical layers $f_s$ and weights $\vec{\Tilde{W}}^S$, quantum layer parameters $\vec{\Tilde{\Theta}}$, Adversary Reconstruction Model $f_\texttt{rec}$, Input Data $\textsf{X}_\texttt{inf}$, Mode Selection $mode$ $(classification/attack)$ 
    
    \textbf{Output:} Classification Output $\hat{\textsf{Y}}_\texttt{inf}$ (if classification mode) or Adversary’s Reconstructed Input $\hat{\textsf{X}}_\texttt{inf}$ (if attack mode)\\
    
    \textbf{Notations:} Client-side smashed data: $\textsf{Z}_\texttt{inf}$, quantum layer expectation output: $\vec{E}_\texttt{inf}$

    \medskip
    \textbf{START OF INFERENCE}
    
    \textbf{Client Side:} \tcp{Forward Pass}
    
    \Indp Compute smashed data: $\textsf{Z}_\texttt{inf} = f_c(\textsf{X}_\texttt{inf}, \Tilde{\vec{W}}^C)$
    
    Sample Laplacian noise: $\eta \sim \text{Laplace}(\mu, b)$ \tcp{Noise sampled from Laplace distribution with mean $\mu$ and scale $b$}
    Add noise to smashed data: $\Tilde{\textsf{Z}}_\texttt{inf} \leftarrow \textsf{Z}_\texttt{inf} + \eta$ \tcp{Obfuscate smashed data}
    
    Send $\Tilde{\textsf{Z}}_\texttt{inf}$ to the server 
    
    \Indm\dotfill{}
    
    \textbf{Server Side:} \tcp{Either Classification or Adversary Attack}
    
    \Indp \textbf{if} $mode = classification$ \textbf{then} 
    
        \Indp Compute expectation values: $\vec{E}_\texttt{inf} = f_q(\Tilde{\textsf{Z}}_\texttt{inf}, \Tilde{\vec{\Theta}})$ \tcp{Quantum Layer} 
        Compute classification output: $\hat{\textsf{Y}}_\texttt{inf} = f_s(\vec{E}_\texttt{inf}, \Tilde{\vec{W}}^S)$ \tcp{Classical Layers} 
        \textbf{return} $\hat{\textsf{Y}}_\texttt{inf}$ \\
    
        \Indm \textbf{else if} $mode = attack$ \textbf{then}
    
        \Indp Attempt to reconstruct input: $\hat{\textsf{X}}_\texttt{inf} \leftarrow f_\texttt{rec}(\Tilde{\textsf{Z}}_\texttt{inf})$ \tcp{Adversary Reconstruction} 
        \textbf{return} $\hat{\textsf{X}}_\texttt{inf}$ 
    
    \Indm\dotfill{}
    
    \textbf{Main Execution:}
    
    \Indp Load input data $\textsf{X}_\texttt{inf}$ and compute smashed data $\textsf{Z}_\texttt{inf}$
    
    Compute $\Tilde{\textsf{Z}}_{inf} \leftarrow$ Sample and add Laplacian noise $\eta$ to $\textsf{Z}_\texttt{inf}$ 
    
    \textbf{if} $mode = classification$ \textbf{then} $\hat{\textsf{Y}}_\texttt{inf} \leftarrow$ Inference($\tilde{\textsf{Z}}_\texttt{inf}$) 
    
    \textbf{else if} $mode = attack$ \textbf{then} $\hat{\textsf{X}}_\texttt{inf} \leftarrow$ Adversary Reconstruction($\Tilde{\textsf{Z}}_\texttt{inf}$) 

    \medskip
    \Indm
    \Indm
    \textbf{END OF INFERENCE}
    
    \Return $\hat{\textsf{Y}}_\texttt{inf}$ (if classification mode) or $\hat{\textsf{X}}_\texttt{inf}$ (if attack mode)
\end{algorithm}
\subsection{\label{subsec: rec_exp} Experiments on Encoding Gate-Based Noise Defense}
In this section, we consider the single-client single-server split learning setup and describe our experiments to test and tune our defense mechanism with varying Laplacian noise layer parameters by (i) comparing the differences between reconstructed ($\hat{\textsf{X}}_\textsf{inf}$) and original ($\textsf{X}_\textsf{inf}$) images in hybrid and classical settings and (ii) comparing the classification performances of HQSL and SL in the presence of the noise layer. From these experiments, we devise the optimal noise parameters that give HQSL a distinct advantage over SL. 
We conduct our experiments on Reconstruction Model 1 to tune our Laplacian noise parameters $\mu$ and $b$. In Appendix \ref{subsec:appendix3}, we present the reconstruction performances of the other two reconstruction models presented in Section \ref{subsec:attack_model} when we apply our proposed noise defense mechanism with the configured parameter values $\mu$ and $b$. 

We split each dataset in the $\textsf{X}_\textsf{train} : \textsf{X}_\textsf{test}$ of 4 : 1 ratio. We train the client and server models in both HQSL and SL settings using the training dataset, $\textsf{X}_\textsf{train}$. The training process takes place as described in Section \ref{subsec: exp_1client_tests}. We further split the test dataset, $\textsf{X}_\textsf{test}$, in the ratio $\textsf{X}_\textsf{rec}$ : $\textsf{X}_\textsf{inf}$ of 4 : 1. $\textsf{X}_\textsf{rec}$ represents a subset of the auxiliary dataset, $\textsf{X}_\textsf{rec}\subset\mathcal{D}_\textsf{aux}$, as described in our threat model (Section \ref{subsubsec: rec_threat_model}), to train the reconstruction model. Here, each dataset is split into three subsets such that they are disjoint from each other. We pass $\textsf{X}_\textsf{rec}$ to the trained client model to generate the smashed data, $\textsf{Z}_\textsf{rec}$. This step assumes that the client model represents the shadow model, $\textsf{f}_\textsf{shadow}$, described in our threat model. We generate the smashed data, $\textsf{Z}_\textsf{rec}$, from the auxiliary dataset and use the data pair ($\textsf{Z}_\textsf{rec}$, $\textsf{X}_\textsf{rec}$) to train the adversary's reconstruction model, $\textsf{f}_\textsf{rec}$.

We use the stochastic gradient descent, \code{SGD}, optimizer, with a learning rate of $10^{-3}$ and momentum of $0.9$, to train the reconstruction model for 200 epochs using the MAE/\code{L1Loss} loss function.  
After training the adversary's reconstruction model, we introduce the Laplacian noise layer, with mean $\mu\in \{0, \pi, 2\pi, 3\pi, 4\pi\}$ and scale $b\in \{0.01, 0.1, 0.5, 1\}$, at the end of the previously trained client-side model.
We consider this range of values for the mean $\mu$ because it represents the location of a qubit at different points along a full rotation around the x-axis of the Bloch sphere. Notably, due to the $4\pi$-periodicity of quantum states under such rotations, angles beyond $4\pi$ do not provide additional insights while tuning our Laplacian noise defence layer, as they result in equivalent quantum states. For example, mean values $\mu=6\pi$ and $8\pi$ would be redundant as they are equivalent to $\mu=2\pi$ and $4\pi$ respectively, as the state evolution repeats cyclically with a period of $4\pi$. The cases of $\mu=2\pi$ and $4\pi$ are particular relevant, as they are expected to cause the quantum states to be invariant in the presence of our noise layer given a small scale $b$ as explained in Section \ref{subsubsec:defence_model}.

Next, $\textsf{X}_\textsf{inf}$ represents a portion of the private data of a client at inference time, $\textsf{X}_\textsf{inf} \subset \mathcal{D}_\textsf{priv}$, that we aim to reconstruct using the trained reconstruction model. However, due to the noise layer, the client model now outputs noisy smashed data, $\tilde{\textsf{Z}}_\textsf{inf}$. $\tilde{\textsf{Z}}_\textsf{inf}$ represents the input to the reconstruction model during the attack phase to recreate $\textsf{X}_\textsf{inf}$ as $\hat{\textsf{X}}_\textsf{inf}$. 

We evaluate the reconstruction performance by computing the difference between the original ($\textsf{X}_\textsf{inf}$) and reconstructed images ($\hat{\textsf{X}}_\textsf{inf}$). For the Fashion-MNIST and MNIST datasets, we use three image comparison metrics: cosine distance, mean square error (MSE), and structural dissimilarity index measure(DSSIM). For the spectrograms from the Speech Commands dataset, the 3 metrics we employ are cosine distance, DSSIM and log-spectral distance (LSD). We used LSD for the Speech Commands spectrograms instead of MSE as LSD gave more significant results than MSE. We describe these metrics in more detail in Appendix \ref{appendix4: metrics}.
To ensure a fair comparison, we apply a masking function to the original and reconstructed images. This masking function takes 2 images of the same dimensions and extracts the non-zero pixel values from both images at positions where at least one image has a non-zero pixel. These masked images are then used for further analysis and comparison using the image comparison metrics.
The means of these metrics are taken over a batch of testing data. We present our findings in Section \ref{subsubsec:metrics}.

For evaluating the classification performance at inference time, we compare the test accuracies and F1-scores under HQSL and SL settings for five folds, now in the presence of the Laplacian noise layer with parameters $\mu$ and $b$. These results are discussed in Section \ref{subsubsec:inference_perf}.
\subsection{\label{subsec: ResultsReconstruction}Reconstruction Attacks Results and Discussion}
In this section, we present and discuss our findings from the experiments carried out to test our proposed encoding gate-based noise defense layer as a countermeasure against reconstruction attacks. We evaluate the performance of the reconstruction model by using the metrics described earlier in \ref{subsubsec:metrics}. This is done in both classical and hybrid settings and for different Laplacian noise parameter values, $\mu$ and $b$.
Next, we analyze HQSL's inference-time accuracies and F1-scores across five folds, fixing the mean parameter $\mu$ while varying the scale parameter $b$, as detailed in Section \ref{subsubsec:inference_perf}.
These findings provide insights into optimizing the noise parameters to enhance HQSL's defense against reconstruction attacks while maintaining robust inference-time classification performance compared to SL, despite the added noise layer.

\subsubsection{\label{subsubsec:metrics}Comparing the Difference between Original and Reconstructed Images under HQSL and SL Settings at Different Noise Levels}
To evaluate the performance of the reconstruction model in classical and hybrid settings in the presence of Laplacian noise, we investigated the impact of noise mean, $\mu$, and scale, $b$, in reconstructing Fashion-MNIST, MNIST, and spectrogram images. Our comparative analysis focused on 4 key image comparison metrics: Mean Cosine Distance, Mean MSE, Mean Structural Dissimilarity Index (Mean DSSIM), and Mean Log Spectral Distance (Mean LSD). The results for the FMNIST, MNIST, and Speech Commands spectrograms datasets are depicted in Fig. \ref{fig:fmnist_metrics_mu}, \ref{fig:mnist_metrics_mu} and \ref{fig:audio_metrics_mu} respectively. The baseline values (represented by the horizontal dashed lines) correspond to the noise-free implementation of the reconstruction attack, i.e., when the smashed data transmitted to the reconstruction model are free from Laplacian noise. In both the classical and hybrid settings, the baseline results are closer to 0 than when the noise layer is applied, showing the reconstruction attacks are successful in recreating the private inputs. These results, hence, underscore the need for the defense mechanism that we investigate in this section.  
\begin{figure*}[htbp!]
    \centering
    \begin{subfigure}{\textwidth}
        \includegraphics[width=\textwidth]{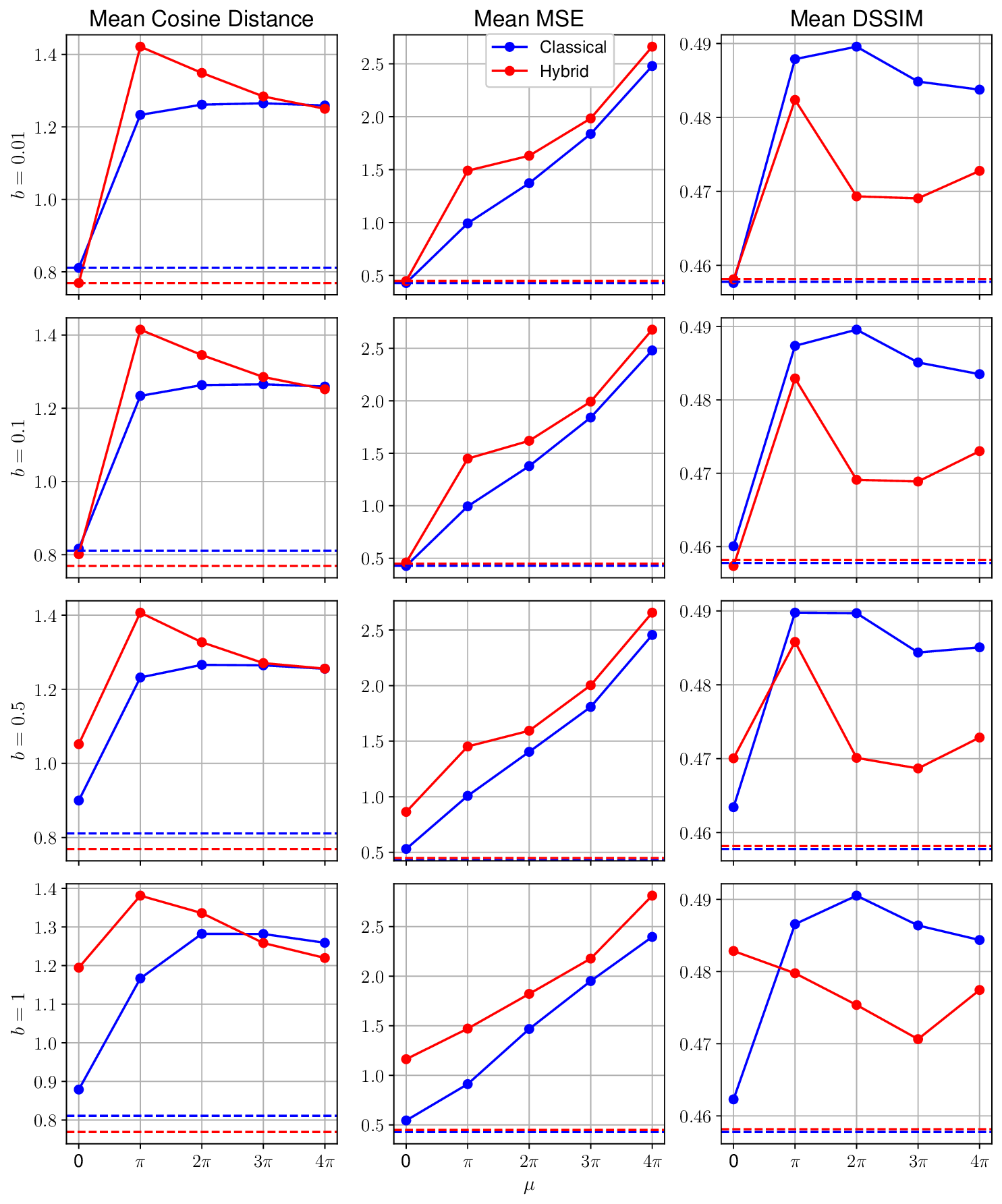}
     \caption{Effect of varying noise parameters $\mu$ and $b$ on FMNIST dataset}
     \label{fig:fmnist_metrics_mu}
     \end{subfigure}
\end{figure*}
\begin{figure*}[htbp!]\ContinuedFloat
    \centering
    \begin{subfigure}{\textwidth}
        \includegraphics[width=\textwidth]{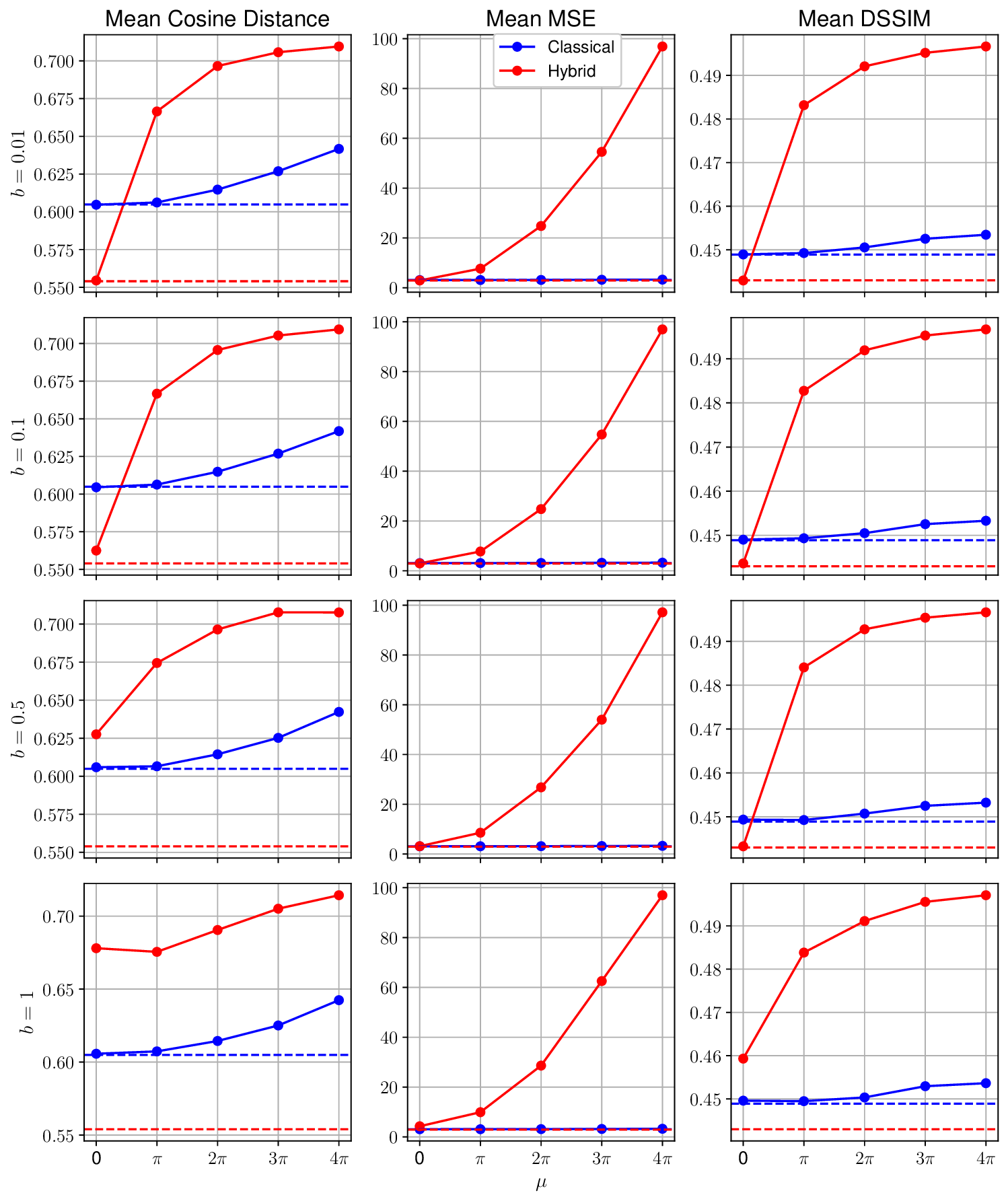}
     \caption{Effect of varying noise parameters $\mu$ and $b$ on MNIST dataset}
     \label{fig:mnist_metrics_mu}
     \end{subfigure}
\end{figure*}
\begin{figure*}[htbp!]\ContinuedFloat
    \centering
    \begin{subfigure}{\textwidth}
    \includegraphics[width=\textwidth]{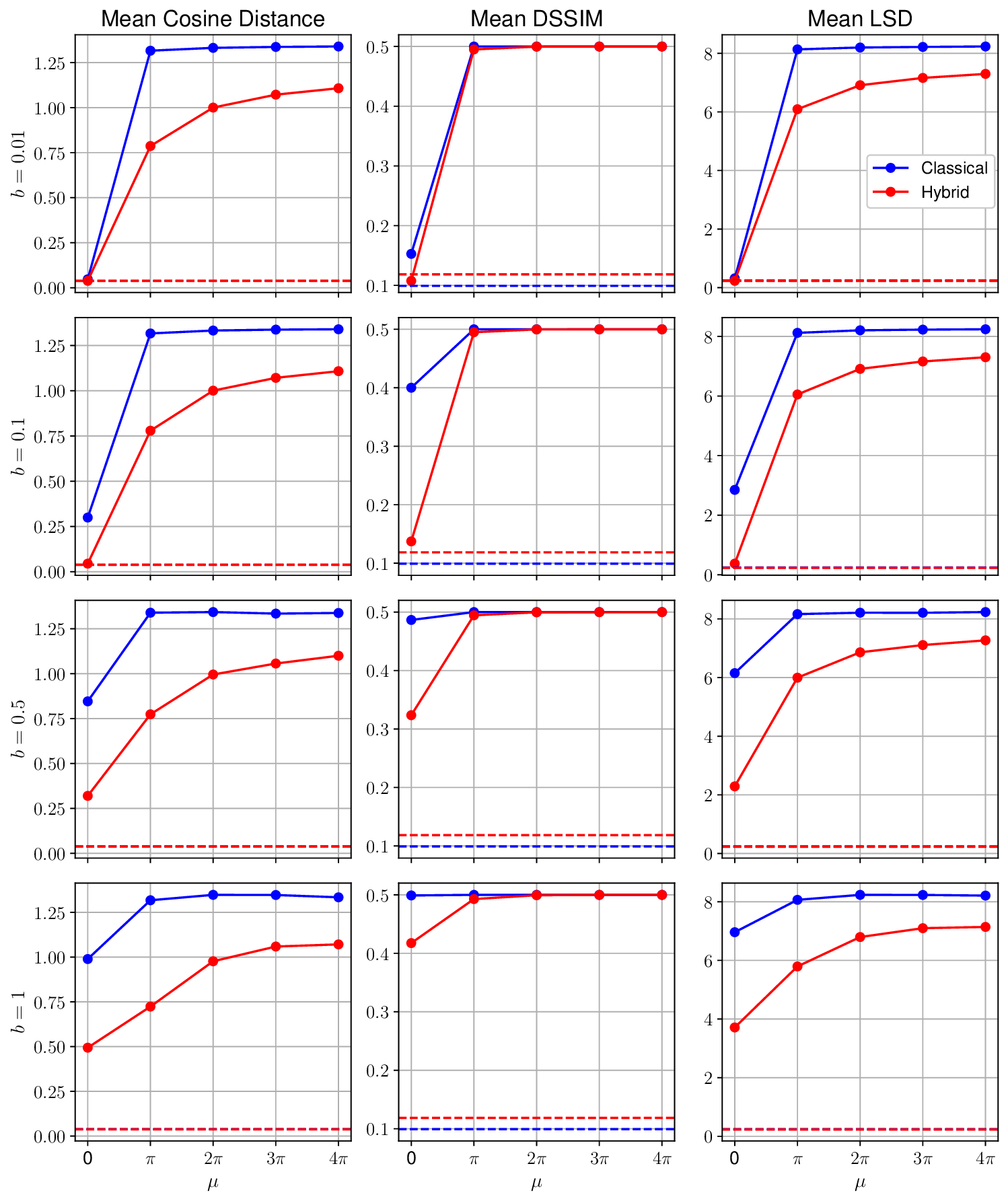}
    \caption{Effect of varying noise parameters $\mu$ and $b$ on Speech Commands spectrograms dataset}
    \label{fig:audio_metrics_mu}
    \end{subfigure}
    \caption{Reconstruction performance with varying Laplacian Noise Layer parameters. The horizontal dashed lines represent the baseline values for each metric, with each line's color corresponding to the legend. Irrespective of the scale parameter value, $b$, at mean $\mu=4\pi$, we obtain, in general, the largest deviations from the baseline. This finding is consistent for all datasets and metrics considered}
    \label{fig:reconstruction_performance}
\end{figure*}

The general trends from Fig. \ref{fig:fmnist_metrics_mu} and \ref{fig:mnist_metrics_mu} were that as we increased the mean parameter, $\mu$, the metric values were well above the baseline, irrespective of the noise scale parameter, $b$. Notably, on the MNIST dataset, the deviations from the baseline in the hybrid setting were more significant than in the classical setting for all 3 metrics. This shows that there was a larger difference between original and reconstructed images in the hybrid setting than in the classical setting. In the classical setting, the minimal deviations from the baseline indicate that a reconstruction attack can still be successful for the MNIST dataset, regardless of the noise parameters. 
This suggests that, at least for the MNIST dataset, the reconstruction model in the classical setting appears relatively insensitive to changes in the noise parameters, with its baseline performance being largely preserved regardless of the trialled noise configurations.
We also highlight that at $\mu=4\pi$, irrespective of the scale parameter value, $b$, the largest deviations were obtained, particularly in the hybrid setting, indicating that at this $\mu$ value, the additive noise was successfully hindering the reconstruction performance of the adversary's model. We also note that significant deviations were still obtained at $\mu=2\pi$. These findings underscore that the mean values $\mu=2\pi$ and $4\pi$ caused the performance of the adversary's reconstruction model to drop in the hybrid setting. 

The results for the Speech Commands spectrograms shown in Fig. \ref{fig:audio_metrics_mu} indicated similar findings. The baseline results were very close to 0, showing that in the absence of the noise layer, the reconstructed spectrograms were very similar to the original spectrograms. As we increased the mean, $\mu$ up to $4\pi$, the deviations increased. For this dataset, the mean cosine distance and mean LSD metrics showed that in the classical setting, the deviations from the baselines were more significant than in the hybrid setting. Still, large deviations above the baseline were obtained in the hybrid setting, signalling the increased difficulty of the adversary's model to recreate the original spectrograms. For all scale parameters, $b$, at $\mu=2\pi$  and $4\pi$, we obtained significant deviations in both the hybrid and classical settings. 

Overall, this study demonstrates that the introduction of the Laplacian noise layer, with appropriately configured noise parameters, can obfuscate the smashed data thereby hindering the performance of the reconstruction model, and in some cases, to a much greater extent in the hybrid setting.
From these results, we establish that mean values of $\mu=2\pi$ and $4\pi$ effectively increase the difference between the original and reconstructed images, hence supporting our analysis in Section \ref{subsubsec:defence_model}. Therefore, we conclude here that \textit{by tuning the Laplacian noise layer parameters considering the rotational properties of encoding gates, we can strengthen the resilience of HQSL against reconstruction attacks. In the classical setup, the reconstruction attack can still successfully recreate private input data.} Next, a necessary investigation is the impact of the perturbation caused by the scale parameter, $b$ on the classification performance of HQSL and SL.

\subsubsection{\label{subsubsec:inference_perf}Inference Time Performance of HQSL and SL At Different Noise Levels}
\begin{figure*}[htpb]
\centering
    \begin{subfigure}{\textwidth}
    \centering
    \includegraphics[width = 0.9\textwidth]{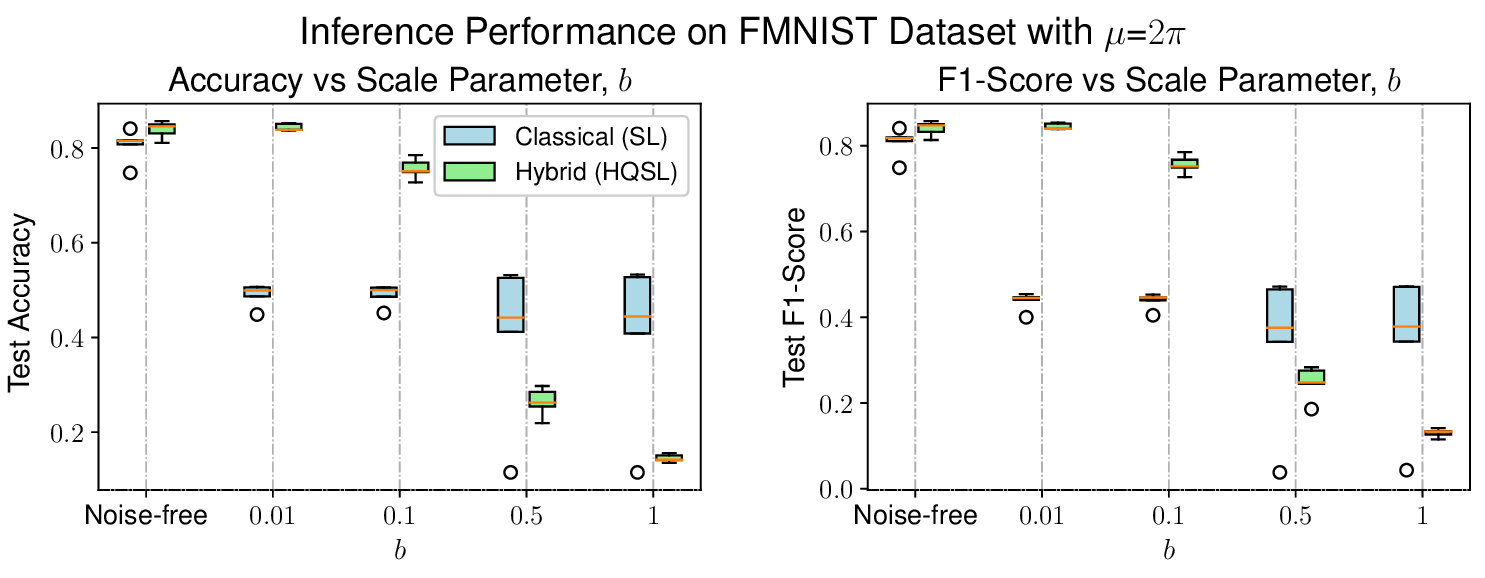}
    \caption{Fashion-MNIST: 5-fold test accuracy and F1-score versus scale parameter, $b$, with mean fixed at $\mu=2\pi$} 
    \label{fig:fmnist_inf_2pi}
    \end{subfigure}
    \begin{subfigure}{\textwidth}
    \centering
    \includegraphics[width = 0.9\textwidth]{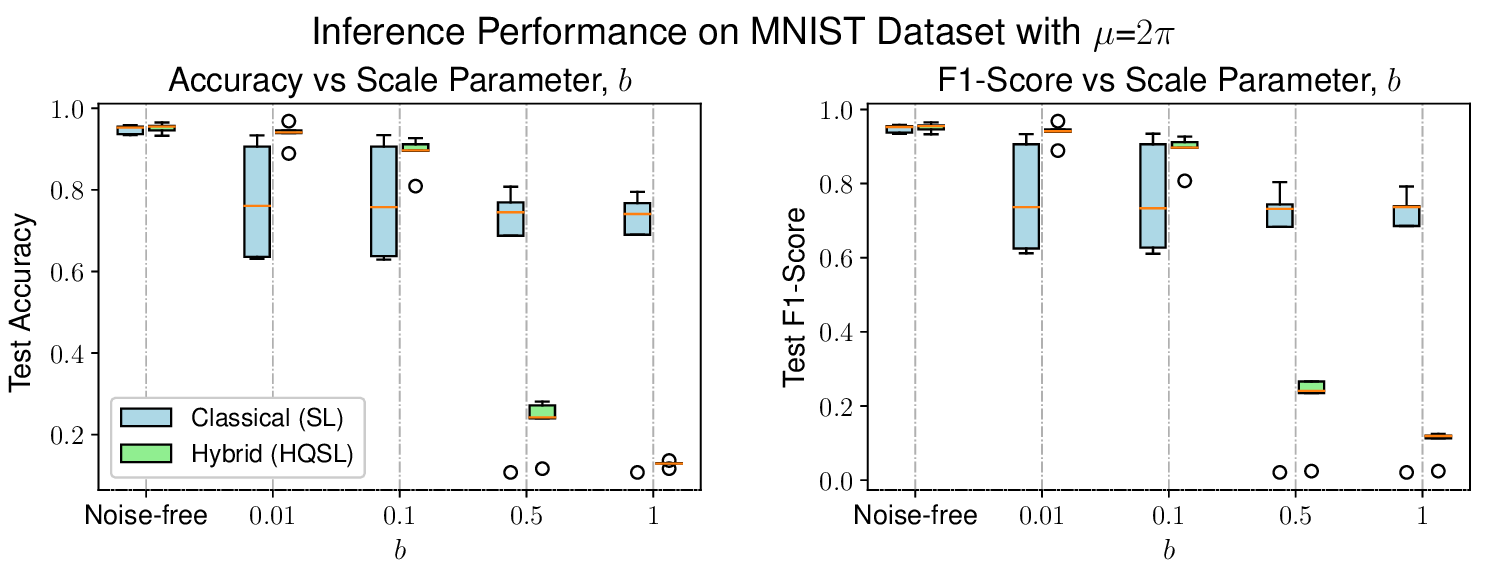}
    \caption{MNIST: 5-fold test accuracy and F1-score versus scale parameter, $b$, with mean fixed at $\mu=2\pi$} 
    \label{fig:mnist_inf_2pi}
    \end{subfigure}
    \begin{subfigure}{\textwidth}
    \centering
    \includegraphics[width = 0.9\textwidth]{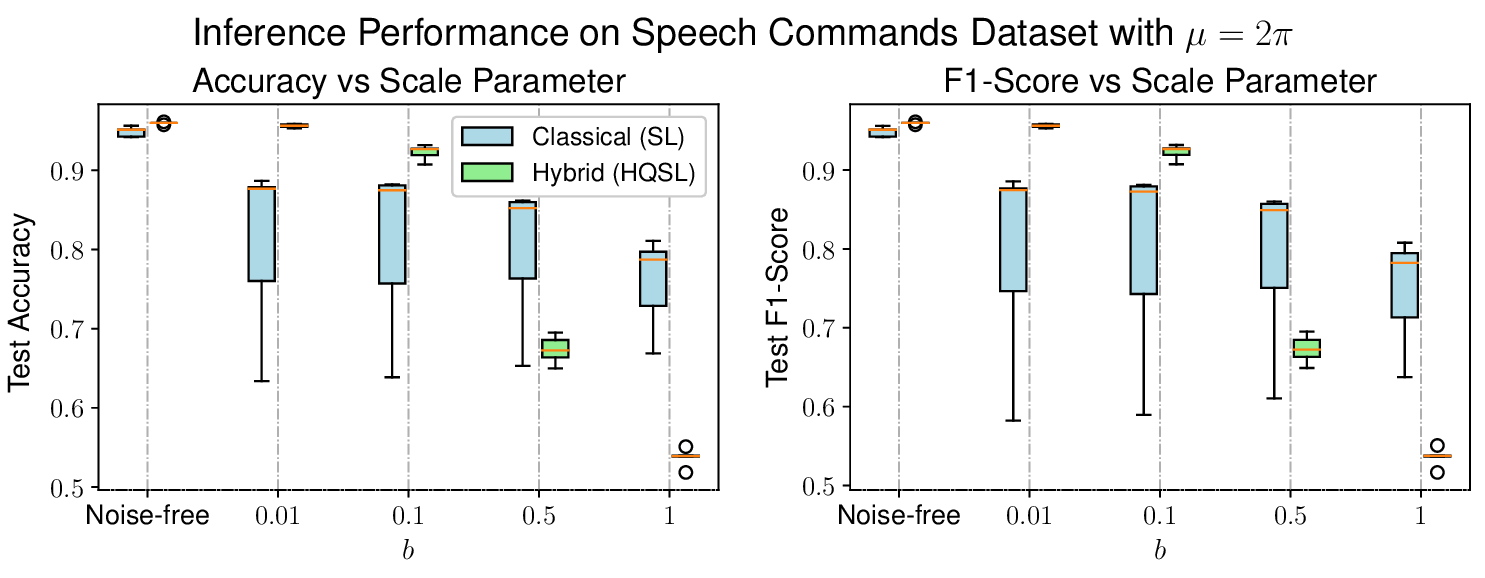}
    \caption{Speech Commands spectrograms dataset: 5-fold test accuracy and F1-score versus scale parameter, $b$, with mean fixed at $\mu=2\pi$} 
    \label{fig:audio_inf_2pi}
    \end{subfigure}
\caption{Effect of noise with fixed mean $\mu=2\pi$ and varying scale $b$ on inference performance in classifying FMNIST, MNIST, and spectrograms from Speech Commands datasets. HQSL (\textcolor{green}{green}) maintains a higher accuracy and F1-score with lower variance than SL (\textcolor{CornflowerBlue}{blue}) in the presence of Laplacian noise with $\mu=2\pi$ and $b = 0.01,0.1$
\label{fig:inference_2pi}
}
\end{figure*}

\begin{figure*}[htpb]
\centering
    \begin{subfigure}{\textwidth}
    \centering
    \includegraphics[width = 0.9\textwidth]{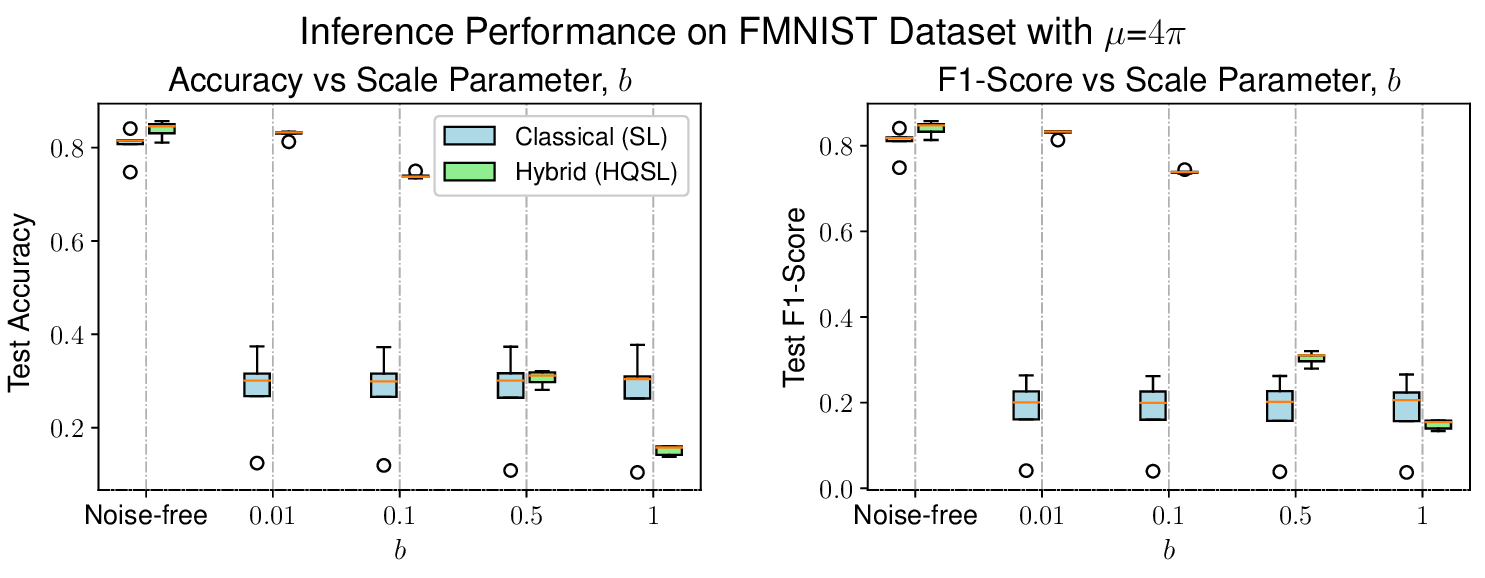}
    \caption{Fashion-MNIST: 5-fold test accuracy and F1-score versus scale parameter, $b$, with mean fixed at $\mu=4\pi$} 
    \label{fig:fmnist_acc_noise}
    \end{subfigure}
    \begin{subfigure}{\textwidth}
    \centering
    \includegraphics[width = 0.9\textwidth]{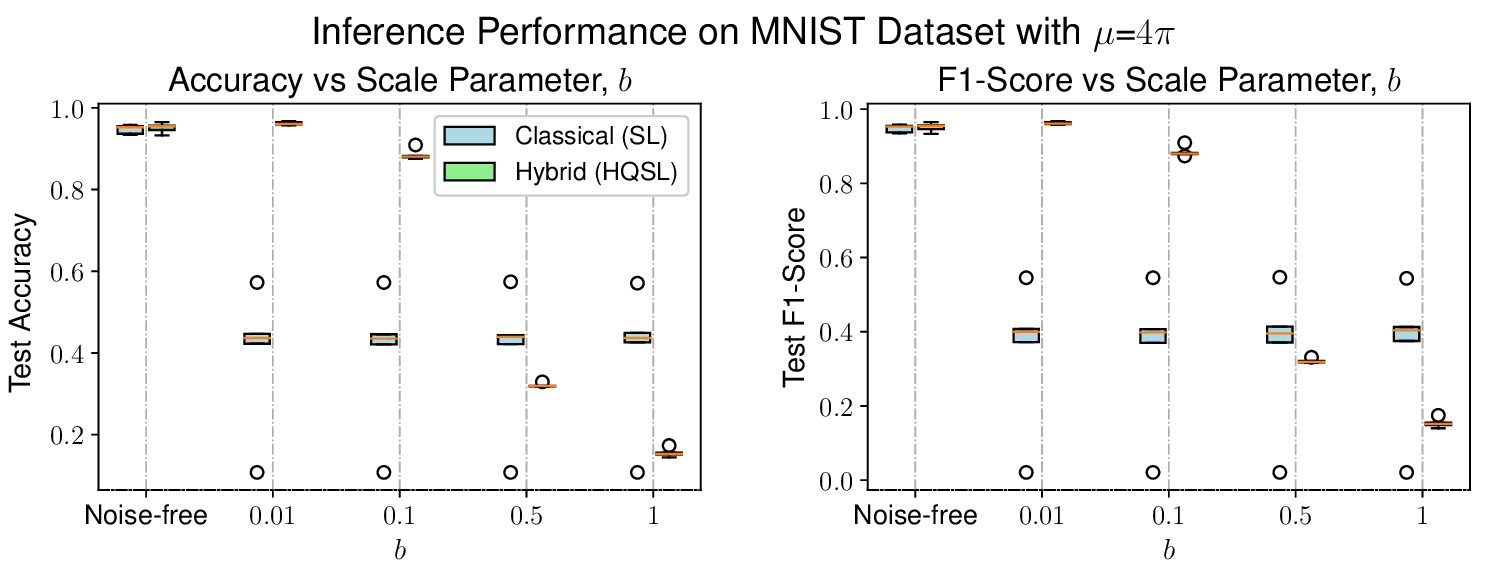}
    \caption{MNIST: 5-fold test accuracy and F1-score versus scale parameter, $b$, with mean fixed at $\mu=4\pi$} 
    \label{fig:mnist_acc_noise}
    \end{subfigure}
    \begin{subfigure}{\textwidth}
    \centering
    \includegraphics[width = 0.9\textwidth]{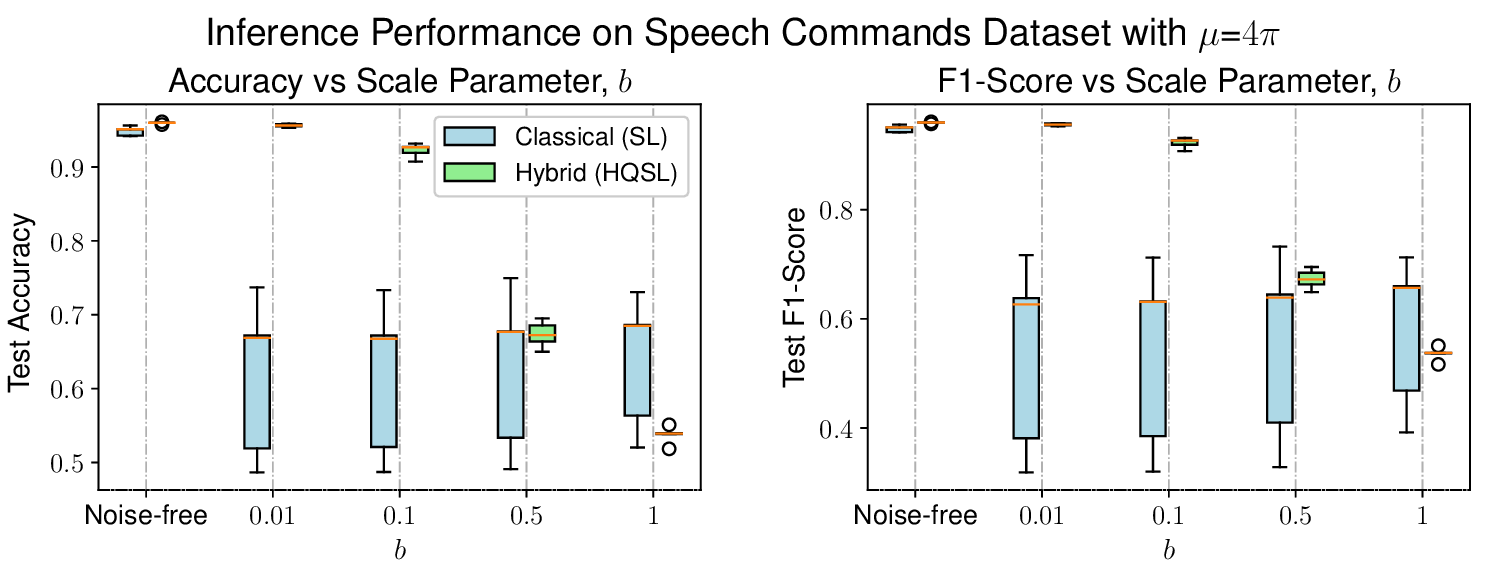}
    \caption{Speech Commands spectrograms dataset: 5-fold test accuracy and F1-score versus scale parameter, $b$, with mean fixed at $\mu=4\pi$} 
    \label{fig:audio_acc_noise}
    \end{subfigure}
\caption{Effect of noise with fixed mean $\mu=4\pi$ and varying scale $b$ on inference performance in classifying FMNIST, MNIST, and spectrograms from Speech Commands datasets. HQSL (\textcolor{green}{green}) maintains a higher accuracy and F1-score with lower fluctuations than SL (\textcolor{CornflowerBlue}{blue}) in the presence of Laplacian noise with $\mu=4\pi$ and $b = 0.01,0.1$
}
\label{fig:noise_inference_performance_4pi}
\end{figure*}

From Section \ref{subsubsec:metrics}, we established that when $\mu=2\pi$ and $4\pi$, we obtain large differences between original and reconstructed images indicated by the deviations above the baseline results. In this section, we investigate the effect of varying scale parameters, $b$, on the inference performance of HQSL and SL given mean values $\mu=2\pi$ and $4\pi$. We measure the performance by comparing the accuracy and F1-score in classifying FMNIST, MNIST, and Speech Commands spectrogram images with varying scale parameters, $b$.
We present these results using the box plots shown in Fig. \ref{fig:inference_2pi} and \ref{fig:noise_inference_performance_4pi}, representing the five-fold test accuracies and F1-scores in the presence and absence of noise. The baseline performances are represented by the Noise-free box plots.

We make the following deductions, backed by the experiments on all 3 datasets. The noise-free box plots representing the baseline performances showed a high accuracy and F1-score with low fluctuations across all folds in both HQSL and SL. The heights of the box plots indicate the fluctuations or variance of the accuracies and F1-score across all folds. By varying the Laplacian noise scale, $b$, and keeping the mean fixed at $\mu=2\pi$ (Fig. \ref{fig:inference_2pi}) and $\mu=4\pi$ (Fig. \ref{fig:noise_inference_performance_4pi}), we demonstrated that a small enough scale parameter value, e.g., $b=0.01$, caused the performance of SL at inference time to drop and show high fluctuations in accuracy and F1-score. In contrast, HQSL maintained a high performance with much less fluctuations -- very similar to the noise-free performance. At both $\mu=2\pi$ and $\mu=4\pi$, when $b=0.1$, HQSL still maintained a superior performance to SL but was slightly worse than when $b=0.01$. This observation shows that when the scale parameter value increased, the performance of HQSL started getting affected by the Laplacian noise layer.
As $b$ increased above 0.1, HQSL's performance at inference time dropped even more. Hence, this experiment indicates that the scale value $b$ should be kept close to 0. We, therefore, choose $b=0.01$ as the desirable scale parameter value for the noise layer as it allows HQSL to maintain a high performance, similar to its noise-free baseline. These experimental findings support our analysis in Section \ref{subsubsec:defence_model}, that the scale parameter value, $b$, should be configured to a small value to limit the perturbations caused by the noise layer, hence, maintaining the HQSL server model's utility. 

Hence, we conclude here that, \textit{by tuning the Laplacian noise layer parameters considering the rotational properties of encoding gates, HQSL's classification performance can match the baseline performance, showing robustness to the noise layer. In contrast, SL is less robust to the tuned noise layer.}

\noindent
\subsubsection*{Reconstruction Performance when HQSL has an Advantage}

\begin{figure}[htpb!]
    \begin{subfigure}{0.5\textwidth}
    \centering
        \includegraphics[width=0.35\textwidth,height=5cm]{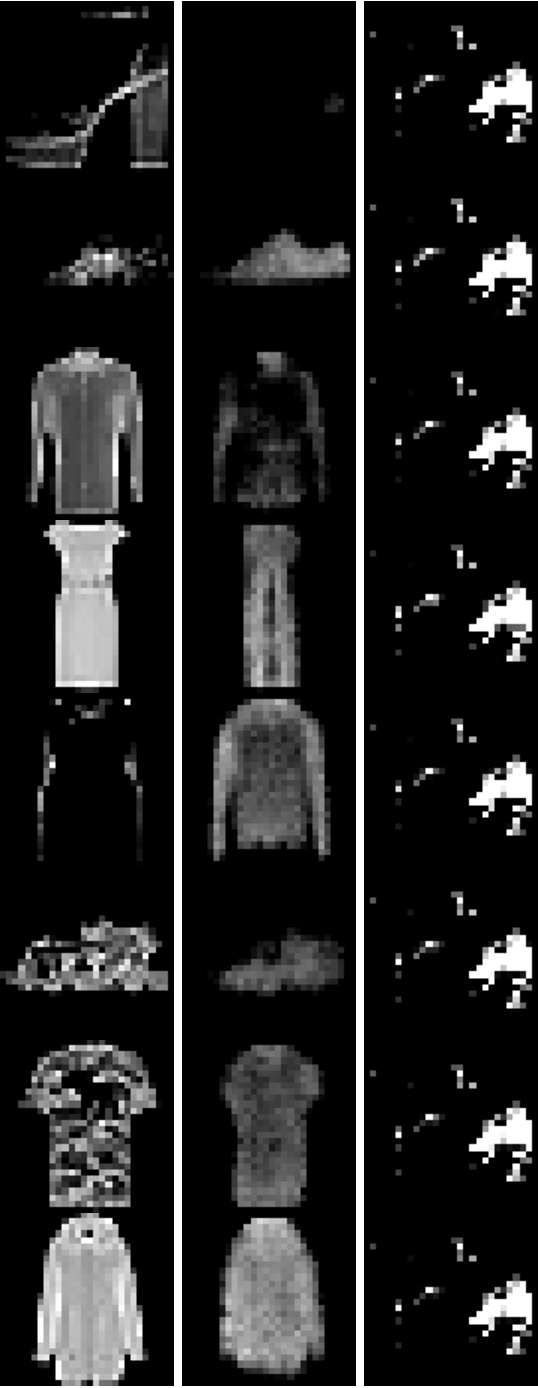}
        \caption{Reconstruction results on FMNIST images}
        \label{fig:rec_FMNIST}
    \end{subfigure}
    \begin{subfigure}{0.5\textwidth}
    \centering
        \includegraphics[width=0.35\textwidth,height=5cm]{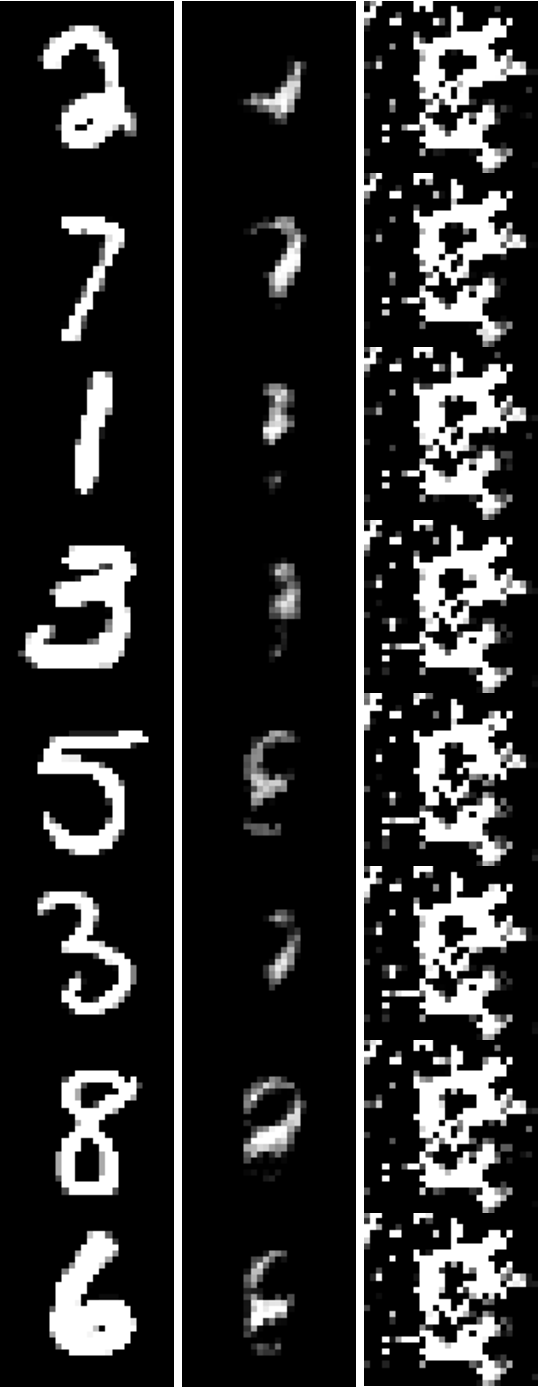}
       \caption{Reconstruction results on MNIST images}
        \label{fig:rec_MNIST}
    \end{subfigure}
    \caption{Left to Right: Original, Reconstructed Images in classical and hybrid settings respectively, with Laplacian noise layer with parameters $\mu=4\pi$, $b=0.01$. For both FMNIST and MNIST images, reconstruction is almost impossible in the hybrid setting, whereas some visual similarities are observed between reconstructed and original images in the classical setting}
    \label{fig:reconstruction_images}
\end{figure}

\noindent
In Fig. \ref{fig:reconstruction_images}, we present some visual observations of the reconstruction attack model performance when we employ our proposed Laplacian noise layer defense tuned considering the rotational properties of quantum encoding gates (here we test using the parameters $\mu=4\pi$ and $b=0.01$). 

Fig. \ref{fig:reconstruction_images} illustrates that the reconstructed images show more significant differences from the original images in the hybrid setting than in the classical setting. Hence, we can directly confirm that our designed noise layer effectively impedes the reconstruction model in recreating the raw images. This is also depicted in the results shown in Fig. \ref{fig:reconstruction_performance}. 
Furthermore, in HQSL, the noise layer causes the smashed data to be handled properly by the quantum layer on the server-side model, hence maintaining a similar classification performance to the noise-free performance as shown in Fig. \ref{fig:inference_2pi} and \ref{fig:noise_inference_performance_4pi}. \textit{This is due to the configured mean values of $\mu=2\pi$ and $4\pi$, which causes the quantum state encoded by the RX-gates in the quantum layer to be nearly invariant to the additive Laplacian noise, given the scale parameter is small ($b=0.01)$.} This means that, for the quantum layer, the obfuscated classical intermediate features correspond to quantum states that are close to those encoded from clean classical smashed data. 

\textit{By leveraging the rotational properties of the encoding gates in the quantum layer of HQSL, we can tune the noise layer parameters to obtain clear advantages in terms of enhancing resistance in split learning against reconstruction attacks from data privacy leakage. The advantages of our proposed noise layer in HQSL are that it (i) impedes the reconstruction attack model's performance in reconstructing raw data, and (ii) enables HQSL to maintain a high performance despite its presence.  HQSL addresses the privacy-utility tradeoff associated with the use of noise layers to enhance resistance against data privacy leakage in SL.}
\section{\label{sec: Conclusion}Conclusion}
This paper addresses the problem of applying Split Learning (SL) concepts to pure Quantum Machine Learning (QML) models for resource-constrained clients lacking quantum computing resources. We propose a novel Hybrid Quantum Split Learning, HQSL, architecture that enables resource-limited clients in the classical domain to train their models collaboratively with a hybrid quantum server for classification tasks. Specifically, HQSL consists of a Hybrid Quantum Neural Network, HQNN, split into two parts: the client side, consisting of a classical neural network, and the server side, comprising an HQNN with the first layer as a quantum layer consisting of a 2-qubit quantum circuit. For the quantum circuit, we introduce a qubit-efficient data-loading technique that allows us to keep the number of qubits and circuit depth small in anticipation of their implementation on near-term quantum devices. 
Experimental results demonstrated the feasibility and ability of HQSL to improve classification accuracy and F1-score compared to its classical equivalent. Performance evaluations on actual quantum devices and noisy simulators provide initial evidence supporting HQSL's noise resilience and practicality for deployment on near-term, noisy quantum (NISQ) devices. Our experiments also demonstrated the scalability of HQSL with an increasing number of clients as HQSL maintains high performance during classification as we sequentially increase the number of clients. 

In this work, we also address the overarching problem of data privacy leakage associated with SL, which can lead to clients' private input data reconstruction attacks on the server side. We propose a Laplacian noise layer defense mechanism designed based on the rotational properties of the encoding gates present within the quantum layer of HQSL. By tuning the parameters of our proposed Laplacian noise layer defense mechanism, we showed that HQSL is more robust to the privacy-utility tradeoff generally associated with the use of noise to enhance SL's resistance against reconstruction attacks in SL.

Overall, HQSL enables resource-constrained clients to train ML models collaboratively with a server equipped with quantum computing resources, resulting in improved classification performance. 
The quantum circuit \rev{designed} in this work underscores a qubit-efficient data-loading mechanism that can be extended beyond the scope of HQSL to address qubit count limitations in near-term devices.
Furthermore, HQSL supports multiple clients in the classical domain for collaborative HQNN model training. 
The defense mechanism proposed in this work enhances HQSL's resistance against data privacy leakage and the risks of reconstruction attacks. 

\noindent
\textbf{Future work.} Although our current HQSL prototype uses a relatively small and shallow quantum circuit, it lays the groundwork for more ambitious and feasible HQNN+SL architectures. As fault‐tolerant devices become available, one could scale HQSL with richer variational blocks or multiple quantum layers on the server side, enabling even lightly resourced clients to benefit from potential quantum advantages. Future studies could also explore alternative split topologies, e.g., U-shaped splits without label sharing or vertically partitioned data, and benchmark HQSL against federated learning (FL) and quantum federated learning (QFL) baselines to provide a more comprehensive performance context. Moreover, while our experiments demonstrate that the rotationally-tuned Laplacian noise layer effectively preserves classification accuracy and impedes reconstruction attacks, a rigorous theoretical analysis, such as formal differential-privacy guarantees or robustness against adaptive adversaries, remains an open area for future research. In addition, future work could investigate defense techniques against a broader range of attack scenarios, beyond reconstruction attacks, e.g., model inversion, membership inference, and quantum-specific attacks, to deepen insights into HQSL’s security vulnerabilities and further strengthen its robustness.

\section*{Declarations}
\begin{itemize}
\item Funding: No funding was received to assist with the preparation of this manuscript.

\end{itemize}

\begin{appendices}
\counterwithout*{figure}{section}
\counterwithout*{table}{section}

\section{\label{sec: appendix-1} Quantum Computing Background}
In this section, we provide further background information on quantum computing supplementing our work. We also discuss the current generation of quantum computers to highlight the regime this paper is relevant to.

\noindent
\textbf{Qubits.} Qubits are the basic unit of information of quantum computing \citep{Nielsen_Chuang_2010}. 
A qubit can be represented mathematically using Dirac notation as $|\psi\rangle=\alpha_0|0\rangle+\alpha_1|1\rangle$, where $|\psi\rangle$ is the quantum state, $\alpha_0, \alpha_1 \in \mathbb{C}$ represent the complex probability amplitudes, and $|0\rangle$ and $|1\rangle$ denote the computational basis states representing a quantum state. This superposition allows for the simultaneous representation of both $|0\rangle$ and $|1\rangle$ states until measured, following the normalization condition $\left|\alpha_0\right|^2 + \left|\alpha_1\right|^2=1$.

\noindent
\textbf{Superposition and entanglement.}
A unique property in quantum mechanics is superposition. 
Superposition allows a quantum state to exist simultaneously in multiple basis states. In quantum computing, this often involves a qubit being in a combination of $|0\rangle$ and $|1\rangle$ states, but superposition can include many more states in more complex systems. 
This property allows calculations to be performed on multiple states at the same time. Entanglement is another property of qubits that occurs when two or more particles become correlated in such a way that their states are no longer independent and cannot be described individually. Instead, their states are intrinsically connected, regardless of the spatial separation between them. 

\noindent
\textbf{Noisy Intermediate-Scale Quantum (NISQ) Computers.}
As predicted by Preskill J. \citep{Preskill2018quantumcomputingin}, we have now entered the NISQ era, where quantum hardware has evolved from a few qubits (e.g., the 2019 IBM Falcon) to modular 
technology boasting 3 x 133 qubits (e.g., the 2024 IBM Heron). However, this generation of hardware is still significantly limited by its size and error rate. It would hence be unrealistic to expect NISQ devices to revolutionize the world on its own; rather, it should be seen as a stepping stone towards developing more advanced quantum technologies in the future. With proper error mitigation and correction techniques, we would be able to run small quantum circuits with desired accuracy for practical applications. \textit{Precisely, in this work, we operate in this regime, where small quantum circuits can have practical applications in QML.}
\section{\label{appendix:datasets}Datasets and Data Processing Details}
In this section, we provide the details of the datasets we use during our experiments, including how we process them. 

\noindent
\textbf{\label{subsubsec: botnetdga}Botnet DGA dataset:}
This dataset was obtained from \citep{IEEEPort}. Botnets consist of compromised devices controlled by a central entity known as the botmaster, and they engage in malicious activities such as launching DDoS\footnote{Distributed Denial-of-Service (DDoS) is an example of unauthorized impairment of electronic communication.} attacks or spreading malware \citep{6489876}. The domain generation algorithm (DGA) is a technique employed by botnets to evade detection and maintain communication with their command-and-control (C\&C) infrastructure. 
For our experiments, we extract 1000 equally distributed samples from the benchmarked dataset, obtained from \citep{codeocean}, consisting of 7 features and 1 binary label (\textit{malicious} or \textit{non-malicious}). 

\noindent
\textbf{\label{subsubsec: breastcancer}Breast Cancer Wisconsin (Diagnostic) dataset:}
A commonly used dataset is the University of California Irvine Machine Learning Breast Cancer Wisconsin (Diagnostic) dataset \citep{misc_breast_cancer_14}. 
This dataset consists of 569 instances of breast cancer samples with 30 attributes. We can classify each sample in this dataset as \textit{benign} or \textit{malignant}. Principal Component Analysis (PCA) is carried out to reduce the number of attributes or features from 30 to only 7 to prevent a high number of trainable parameters. 

\noindent
\textbf{\label{subsubsec:mnist}MNIST dataset:}
This widely used dataset consists of 70,000 28 x 28 grayscale images of handwritten digits (0, \ldots, 9) and their respective labels. They were obtained from \citep{deng2012mnist} and made available for our use using the \code{torchvision} library. For our experiments, we utilized 6000 normalized random samples using a set random seed.

\noindent
\textbf{\label{subsubsec:fmnist}Fashion-MNIST dataset:}
This is another image dataset consisting of 70,000 examples of clothing items and 10 labels. Each sample is a 28 x 28 grayscale image associated with a label from the 10 classes. We obtained the Fashion-MNIST (or FMNIST) dataset from \citep{xiao2017fashionmnist}, extracted 6,000 random samples using the same set random seed as in the MNIST case, and normalized the images.

\noindent
\textbf{\label{subsubsec:speechcommand}Speech Commands dataset:}
This is an audio dataset consisting of 105,829 utterances of 35 commonly spoken words from 2168 speakers \citep{warden2018speech}. Each sample is stored as a \code{.wav} file with a length of a maximum of one second. This dataset is obtained from the \code{torchaudio} library. We extract the utterances corresponding to the words ``yes" and ``no" and generate their respective spectrograms resized into single channel 28 x 28 images. We transform the spectrograms as such to be able to use a similar HQSL model as variant 2 as described in Section \ref{subsubsec:HQSL_modelling}. Hence, in this work, the classification of the Speech Commands dataset is a binary image classification task, where the images are of the same shape as MNIST and FMNIST datasets. 
\section{\label{subsec:appendix1}Using a Heuristic Method to select Quantum Circuit}
In this section, we describe the heuristic approach we have employed to select the quantum circuit that we use to build the quantum layer for its introduction in HQSL.
We adopt this approach to find the optimal quantum circuit by making informed decisions and reaching a solution that satisfies some metric. We state the metric as follows:
\textit{Which quantum circuit from our proposed set of quantum circuits can give rise to an HQSL model that outperforms SL in terms of test accuracy and F1-score for all datasets?} For the classical counterpart (SL) of each of these circuits, we follow the same rule as introduced earlier, i.e., we use a hidden layer where the number of input nodes is equal to the size of the input data dimension, the number of output nodes is the same as the number of qubits (or measurement outputs) in the circuit and the \code{ReLU} activation at the end of the dense layer.

Two important considerations for our set of quantum circuits are the circuit depth and width (number of qubits). We want to have the smallest circuit that satisfies the above-mentioned metric. The compact size of the circuit is advantageous to ensure a reasonable runtime for our simulations using the \code{default-qubit} simulator by PennyLane. Moreover, this makes it more likely to be implemented on real devices in the near-term. In short, the larger the number of qubits and the deeper the circuit, the longer the runtime of HQSL and the harder it is to implement on actual quantum computers.

First, we investigated 2-dimensional inputs to a quantum circuit with 2 qubits and entanglement, where we uploaded each feature 3 times per qubit. This is Circuit 1. In Circuit 2, we scaled the number of qubits to 4, to accommodate 4-dimensional inputs and updated the entanglement strategy accordingly.
In Circuit 3, we kept the 4 qubits layout but adopted a traditional VQC configuration instead with phase-encoding to convert 4-dimensional classical data to quantum states, that were operated on by 3 parameterized rotation gates each and entangling gates. 
We then considered assessing the performance without entangling gates in Circuit 4, which was similar to Circuit 1 but lacked entanglement. In Circuit 5, we increased the depth of the circuit. 
For all the circuits we described above (1-5) when used in HQSL, we were not able to obtain an accuracy and F1-score that outperformed their equivalent SL model on all datasets. Hence, our set metric was not satisfied for these circuits.
We then adapted the work done by \citep{perez-salinas_data_2020} to develop Circuit 6 as described in Section \ref{subsubsec: qlayer}. Our experiments showed that Circuit 6 satisfies the metric we have set and hence, concludes this heuristic approach.

We display all the circuits we have discussed above in Table \ref{tab:circuits_trialled}. A brief description of the quantum circuits is also provided. 
\begin{table*}[htbp!]
  \centering
  \caption{Table of Circuits Trialled and Brief Description of the Circuits}
  \label{tab:circuits_trialled}
  \begin{tabular}{| c | l | p{6cm} |}
    \hline
     \multirow{2}{*}{ID} & \multicolumn{1}{c|}{\multirow{2}{*}{Quantum Circuit}} & \multirow{2}{*}{Description}\\ 
     & & \\
     \hline
    1 & \begin{minipage}{.4\textwidth}\label{fig:circuit_1}
    \vspace{2ex}
      \includegraphics[width=\linewidth]{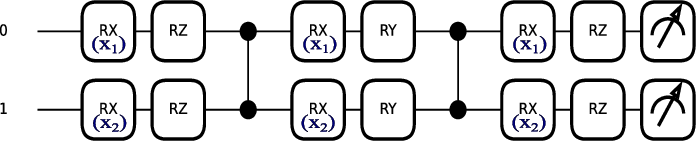}
    \end{minipage}
    & \vspace{-5ex}2-qubit circuit with CZ-entanglement operating on 2 classical input features $(x_1,x_2)$, each uploaded 3 times at the 3 $R_x$-gates on each qubit
    \\ \hline
        2\label{fig:circuit_2} & \begin{minipage}{.4\textwidth}
        \vspace{2ex}
      \includegraphics[width=\linewidth]{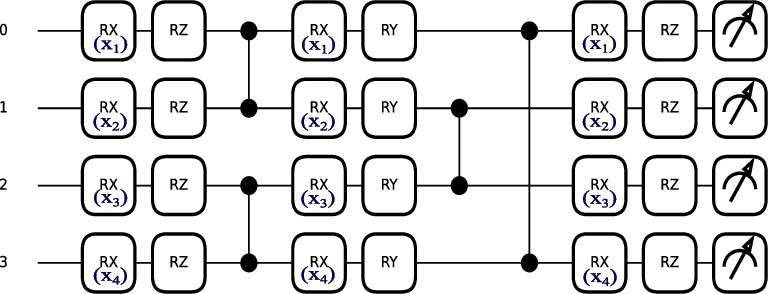}\\
    \end{minipage}
    &\vspace{-8ex}4-qubit circuit with CZ-entanglement organized as shown operating on 4 classical input features $(x_1,x_2,x_3,x_4)$, each uploaded at 3 different points at the RX-gates on each of the 4 qubits
    \\ \hline
        3\label{fig:circuit_3} & \begin{minipage}{.4\textwidth}
        \vspace{2ex}
      \includegraphics[width=\linewidth]{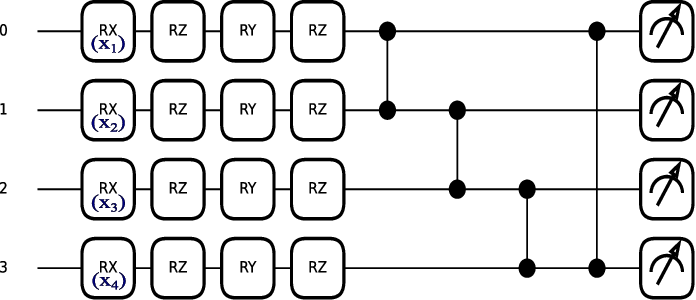}
    \end{minipage}
    &\vspace{-8ex}4-qubit circuit with single-upload of each of the 4 classical input features $(x_1,x_2,x_3,x_4)$ at the encoding layer, followed by the RZ-RY-RZ parameterized gates on each qubit. This is then followed by circular entanglement using CZ-gates. This circuit is an example of a traditional Variational Quantum Circuit.
    \\ \hline
        4\label{fig:circuit_4} & \begin{minipage}{.4\textwidth}
        \vspace{2ex}
      \includegraphics[width=\linewidth]{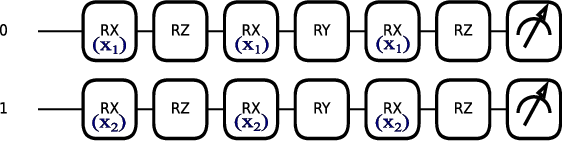}
    \end{minipage}
    &\vspace{-5ex}2-qubit circuit without entanglement operating on 2 classical input features $(x_1,x_2)$, each uploaded 3 times at the RX-gates on each qubit
    \\ \hline
        5\label{fig:circuit_5} & \begin{minipage}{.4\textwidth}
        \vspace{2ex}
      \includegraphics[width=\linewidth]{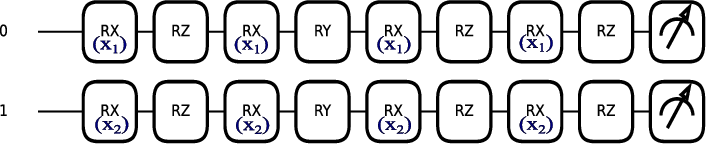}
    \end{minipage}
    &\vspace{-5ex}2-qubit circuit without entanglement operating on 2 classical input features $(x_1,x_2)$, each uploaded 4 times at the RX-gates on each qubit
    \\ \hline 
        6\label{fig:circuit_6} & \begin{minipage}{.4\textwidth}
        \vspace{2ex}
      \includegraphics[width=\linewidth]{Fig3.eps}
    \end{minipage}
    &\vspace{-5ex}2-qubit circuit with CZ-entanglement operating on 3 classical input features $(x_1,x_2,x_3)$, each uploaded once on each qubit at the RX-gates
    \\ \hline
    \end{tabular}
\begin{tikzpicture}[remember picture,overlay]
\foreach \Val in {up,down}
{
\draw[rounded corners,red,thick]
  ([shift={(-0.5\tabcolsep,-36ex)}]pic cs:start\Val) 
    rectangle 
  ([shift={(-62\tabcolsep,-47.5ex)}]pic cs:end\Val);
}
\end{tikzpicture}
\end{table*}
We train and test HQSL on all datasets and compare the testing accuracy and F1-score against their classical counterparts. We present these results in Table \ref{tab:net_results}. The greyed-out areas demonstrate that these experiments were not carried out as the circuits they corresponded to were already rejected as they failed the set metric.
\begin{table*}[htbp!] 
\centering
\caption{Summary of accuracy and F1-score for HQSL model using each quantum circuit from Table \ref{tab:circuits_trialled} compared against their equivalent SL models. Only HQSL with Circuit 6 in its quantum layer outperforms SL on all datasets considered}
\label{tab:net_results}
\resizebox{\columnwidth}{!}{%
\begin{tabular}{ccccccccc}

\hline Datasets & \multicolumn{2}{c}{} & Circuit 1& Circuit 2& Circuit 3& Circuit 4& Circuit 5& Circuit 6\\
\hline \multirow{4}{*}{ Botnet DGA } & \multirow{2}{*}{ Hybrid } & Mean Accuracy & 0.921 & 0.922&\cellcolor{green}0.930&0.918&0.920&\cellcolor{green}0.920 \\
\cmidrule{3-9} & & Mean F1-Score & 0.893 & 0.895 &\cellcolor{green}0.908&0.891&0.890&\cellcolor{green}0.894\\
\cmidrule{2-9} & \multirow{2}{*}{ Classical } & Mean Accuracy & \cellcolor{green}0.928 & \cellcolor{green}0.933&0.923	&\cellcolor{green}0.928&	\cellcolor{green}0.929&	0.919
 \\
\cmidrule{3-9} & & Mean F1-Score & \cellcolor{green}0.904 & \cellcolor{green}0.909&0.897&	\cellcolor{green}0.904&	\cellcolor{green}0.905&	0.890
 \\
\midrule \midrule\multirow{4}{*}{ Breast Cancer } & \multirow{2}{*}{ Hybrid } & Mean Accuracy & 0.958 & \cellcolor{green}0.967 &0.961	&\cellcolor{green}0.970	&\cellcolor{green}0.972&	\cellcolor{green}0.970
\\
\cmidrule{3-9} & & Mean F1-Score & 0.944 & \cellcolor{green}0.956 &0.949&	\cellcolor{green}0.960	&\cellcolor{green}0.962	&\cellcolor{green}0.961
\\
\cmidrule{2-9} & \multirow{2}{*}{ Classical } & Mean Accuracy & \cellcolor{green}0.968 & 0.963 &\cellcolor{green}0.968&	0.968	&0.970&	0.965
\\
\cmidrule{3-9} & & Mean F1-Score & \cellcolor{green}0.959 & 0.951 &\cellcolor{green}0.958&	0.959	&0.960&	0.953\\
\hline \hline \multirow{4}{*}{ FMNIST } & \multirow{2}{*}{ Hybrid } & Mean Accuracy & \cellcolor{green}0.830 & \cellcolor{green}0.843 &0.840&	\cellcolor{green}0.835	& \cellcolor{gray}&	\cellcolor{green}0.839

\\
\cmidrule{3-9} & & Mean F1-Score & \cellcolor{green}0.831 & 0.837 &0.842&	\cellcolor{green}0.838 &\cellcolor{gray}	& \cellcolor{green}0.840
\\
\cmidrule{2-9} & \multirow{2}{*}{ Classical } & Mean Accuracy & 0.809 & 0.838&\cellcolor{green}0.846	&0.809& \cellcolor{gray}&	0.805
\\
\cmidrule{3-9} & & Mean F1-Score & 0.811 & \cellcolor{green}0.840&\cellcolor{green}0.846&	0.811	& \cellcolor{gray}&	0.808  \\
\midrule \midrule \multirow{4}{*}{ MNIST } & \multirow{2}{*}{ Hybrid } & Mean Accuracy & \cellcolor{green}0.953 & \cellcolor{gray}&\cellcolor{green}0.965&	0.952	&\cellcolor{gray} &	\cellcolor{green}0.951
\\
\cmidrule{3-9} & & Mean F1-Score & \cellcolor{green}0.953 &\cellcolor{gray} &\cellcolor{green}0.965&	0.952& \cellcolor{gray}&	\cellcolor{green}0.951
\\
\cmidrule{2-9} & \multirow{2}{*}{ Classical } & Mean Accuracy & 0.952 &\cellcolor{gray} & 0.964&	0.952	& \cellcolor{gray}&	0.948
\\
\cmidrule{3-9} & & Mean F1-Score & 0.952 & \cellcolor{gray}&0.964&	0.952	& \cellcolor{gray}&	0.948\\
\midrule \midrule \multirow{4}{*}{ Speech Commands } & \multirow{2}{*}{ Hybrid } & Mean Accuracy &\cellcolor{gray} & \cellcolor{gray}&\cellcolor{gray}&	\cellcolor{gray}	&\cellcolor{gray} &	\cellcolor{green}0.960 
\\
\cmidrule{3-9} & & Mean F1-Score & \cellcolor{gray} &\cellcolor{gray} &\cellcolor{gray}&	\cellcolor{gray}& \cellcolor{gray}&	\cellcolor{green}0.960
\\
\cmidrule{2-9} & \multirow{2}{*}{ Classical } & Mean Accuracy &\cellcolor{gray} &\cellcolor{gray} &\cellcolor{gray}&	\cellcolor{gray}	& \cellcolor{gray}&0.949
\\
\cmidrule{3-9} & & Mean F1-Score & \cellcolor{gray} & \cellcolor{gray}&\cellcolor{gray}&	\cellcolor{gray}	& \cellcolor{gray}&	0.948\\
\bottomrule
\end{tabular}
}

\begin{tikzpicture}[remember picture,overlay]
\foreach \Val in {up,down}
{
\draw[rounded corners,red,thick]
  ([shift={(30.5\tabcolsep,56ex)}]pic cs:start\Val) 
    rectangle 
  ([shift={(24.9\tabcolsep,0ex)}]pic cs:end\Val);
}
\end{tikzpicture}
\end{table*}

As shown in Table \ref{tab:net_results}, only Circuit 6 gave an outperformance in terms of both mean accuracy and F1-score over the equivalent SL model for all datasets studied. Circuit 6 is the one that makes efficient use of the 2 qubits in the circuit, whereby we load each of the 3 classical input features onto each qubit at 3 different RX-gates. Hence, we chose this qubit-efficient data-loading quantum circuit due to its superior performance compared to its classical analogue and, it is also easy to simulate without significant computational complexity.

\section{\label{appendix4: metrics}Metrics to compare Original and Reconstructed Images}
We describe the metrics that we use to compare the difference between the client's original input and reconstructed images.
\theoremstyle{definition}
\newtheorem{definition}{Metric}
\begin{definition}
\textbf{Cosine Distance, $\textbf{D}_\textbf{c}$:} $D_c$ is a metric used to complement cosine similarity. The latter is a measure of similarity between two non-zero vectors and is defined as the cosine of the angle between the vectors. While cosine similarity metric has been used in multiple contexts, such as face verification \citep{10.1007/978-3-642-19309-5_55} and text analysis \citep{7577578}, we use cosine distance to indicate the difference between two vectors instead of their similarity.
Hence, the larger the cosine distance is, the larger the difference between two images, $\vec{A}$ and $\vec{B}$.
\begin{equation}
    \text{Cosine Similarity} , S_c(\vec{A},\vec{B}) \coloneqq \frac{\vec{A}\cdot\vec{B}}{||\vec{A}||\hspace{2pt}||\vec{B}||} ,-1 \leq S_c \leq 1
\end{equation}
\begin{equation}
    \text{Cosine Distance}, D_c(\vec{A},\vec{B}) = 1-S_c(\vec{A},\vec{B}),    0 \leq D_c \leq 2
\end{equation}
\end{definition}

\begin{definition}
    \textbf{Mean Square Error, MSE}: MSE is a commonly used metric to compare the difference between 2 images. We use it to compare the difference between the reconstructed images and the original images. The greater the MSE is, the greater the difference between the 2 images. For two images $\vec{A}$ and $\vec{B}$ of size M×N, the MSE is calculated as:
\begin{equation}
\mathrm{MSE}=\frac{1}{MN} \sum_{i=1}^{M} \sum_{j=1}^{N} (\vec{A}(i,j) - \vec{B}(i,j))^2
\end{equation}
where $A(i,j)$ and $B(i,j)$ are the pixel values of the original and reconstructed images, respectively, at position $(i,j)$, and M and N are the dimensions of the images.
\end{definition}

\begin{definition}
    \textbf{Structural Dissimilarity Index, DSSIM}: DSSIM can be considered as the complement of the structural similarity index, SSIM. The SSIM metric is widely used to quantify the quality of images and measure the similarity between 2 images \citep{wang2004image}. The SSIM index is a real number between -1 and 1, where 1 indicates perfect similarity, 0 indicates no similarity, and -1 indicates perfect anti-correlation. For our purpose, we treat values less than 0 as showing no similarity with the original image. Hence, we re-define SSIM as $\text{SSIM} = max(\text{SSIM}, 0)$ such that now, $0 < \text{SSIM} < 1$.
    While we do not delve into the mathematical properties of SSIM, we define DSSIM as another measure of the difference between 2 images. 
\begin{equation}
    \text{DSSIM}(\vec{A}, \vec{B}) = \frac{(1-\text{SSIM}(\vec{A}, \vec{B}))}{2}, 0\leq DSSIM\leq 0.5
\end{equation}
Given our redefinition of SSIM, similar to the previous two metrics, the larger the DSSIM value is, i.e., the closer it is to 0.5, the larger the difference between the reconstructed and original images.
\begin{definition}
    \textbf{Log Spectral Distance, LSD:} LSD is another metric that has been used in \citep{nugraha2020flow} that can be used to compare the original and reconstructed spectrograms. LSD has been used in speech recognition comparing the power spectra of two speech signals \citep{115972}. We apply this to the original and reconstructed spectrograms from the speech commands dataset. The LSD between 2 images $\vec{A}$ and $\vec{B}$ of size M×N can be computed as
    \begin{equation}
        \text{LSD} = \sqrt{\frac{1}{MN} \sum_{i=1}^{M} \sum_{j=1}^{N}\left[\log_{10} \left( \frac{\vec{A}(i,j)}{\vec{B}(i,j)} \right) \right]^2}
    \end{equation}
\end{definition}
\end{definition} 
\section{\label{subsec:appendix3}Additional Experiments with Reconstruction Attack Models 2 and 3}
By carrying out experiments on Reconstruction Model 1, we have seen that when the noise parameters are tuned to $\mu=2\pi, 4\pi$ and $b=0.01$, our HQSL model is more resilient to reconstruction attack by the adversary's model. In this section, we show that similar resilience is shown against other reconstruction models obtained from the literature. We described these models in Section \ref{subsec:attack_model}. We configure our noise layer with the parameters $\mu=4\pi$ and $b=0.01$ in testing the reconstruction models against our encoding gate-based defense mechanism.
These models form the decoder part of an auto-encoder and were obtained from \citep{li2018discriminatively} -- Reconstruction Model 2 and \citep{balle2022reconstructing} -- Reconstruction Model 3.  To demonstrate the reconstruction performance in the hybrid and classical settings, we again use the metrics described in Appendix \ref{appendix4: metrics}. The results are presented in Tables \ref{tab_appen:reconstructionresults} - Fashion-MNIST and MNIST reconstructions and \ref{tab_appen:reconstructionresults_audio} - Speech Commands spectrogram reconstruction.
\begin{table*}[htbp!]
\centering
\caption{Comparison of Mean Cosine Distance, MSE and MDSSIM values between raw Fashion-MNIST (FMNIST) and MNIST images and reconstructed images for reconstruction models 2 and 3, with noise parameters $\mu=4\pi, b=0.01$. The larger the metric, the worse the reconstruction performance. The values in the bracket represent the baseline results, corresponding to noise-free reconstruction performance}
\label{tab_appen:reconstructionresults}
\resizebox{\columnwidth}{!}{%
\begin{tabular}{|c|c||c|c||c|c||c|c|}
\hline 
&  Reconstruction& \multicolumn{2}{c||}{ \underline{Mean Cosine Distance}} & \multicolumn{2}{c||}{ \underline{MSE} } & \multicolumn{2}{c|}{ \underline{MDSSIM} } \\
& Models& Classical & Hybrid & Classical & Hybrid & Classical & Hybrid \\
 \hline \multirow{2}{*}{ FMNIST } & Model 2 & 1.211 (0.539)& \textbf{1.253(0.559)}& 1.341(0.253)& \textbf{16.972(0.248)}& 0.484(0.443)& \textbf{0.496(0.455)}\\
& Model 3 & 1.000(0.998)& \textbf{1.377(0.638)}& 0.302(0.302)& \textbf{8.475(0.805)}& 0.489(0.489)& \textbf{0.499(0.499)}\\

\hline 
\multirow{2}{*}{ MNIST } & Model 2  & 0.992(0.674)& \textbf{1.379(0.675)}& 1.003(0.587)& \textbf{11.694(0.598)}& 0.480(0.457)& \textbf{0.494(0.462)}\\
                         & Model 3 & 1.000(1.000)& \textbf{1.219(0.814)}& 0.699(0.699)& \textbf{16.923(0.649)}& 0.483(0.474)& \textbf{0.499 (0.499)}\\

\hline
\end{tabular} 
}
\end{table*}
\begin{table*}[htbp!]
\centering
\caption{Comparison of Mean Cosine Distance, MDSSIM and Mean Log Spectral Distance (Mean LSD) values between original and reconstructed spectrogram images from Speech Commands dataset for reconstruction models 2 and 3, with noise parameters $\mu=4\pi, b=0.01$. The larger the metric, the worse the reconstruction performance. The bracketed values represent the baseline metric values, corresponding to noise-free performance}
\label{tab_appen:reconstructionresults_audio}
\resizebox{\columnwidth}{!}{%
\begin{tabular}{|c|c||c|c||c|c||c|c|}
\hline 
&  Reconstruction& \multicolumn{2}{c||}{ \underline{Mean Cosine Distance}} & \multicolumn{2}{c||}{\underline{MDSSIM}} & \multicolumn{2}{c|}{ \underline{Mean LSD} } \\
&  Models & Classical & Hybrid & Classical & Hybrid & Classical & Hybrid \\
 \hline Speech & Model 2 & \textbf{0.333(0.024)}& 0.179(0.014)& \textbf{0.479(0.149)}& 0.403(0.094)& \textbf{5.544(1.820)}& 2.932(1.379)\\
Commands & Model 3 & 0.791(0.031)& \textbf{0.955(0.006)}& 0.497(0.190)& \textbf{0.499(0.075)}& 9.303(2.152)& \textbf{9.629(0.597)}\\
\hline 
\end{tabular} 
}
\end{table*}

In general, the reconstruction performance is worse in the hybrid settings, showing reconstruction becomes harder with the introduction of Laplacian noise with $\mu=4\pi$ and $b=0.01$. Significant deviations from the baselines are obtained in most cases. When coupled with the inference performance comparison from Fig. \ref{fig:noise_inference_performance_4pi}, we show a clear advantage of HQSL compared to SL. Specifically, in the presence of our proposed noise defense mechanism based on the rotational properties of the encoding gates in the quantum layer, we can simultaneously impede the reconstruction performance in the hybrid setting and maintain a high classification performance by HQSL. In contrast, SL's performance during classification suffers from the defense method we employ.

\section{\label{appendix:speech}Speech Commands Dataset Result for Single Client-Single Server Experiments}
In this section, we present the raw data collected for the Speech Commands dataset describing the accuracies and F1-scores obtained during the single client-single server experiments. The results presented in Table \ref{tab:performance_metrics_speech} complement the single client-single server results in Fig. \ref{fig:feasibility} presented in Section \ref{subsec: exp_1client_tests}. The data show that the extremely low standard deviations (std) in the accuracies and F1-scores for HQSL demonstrate that our HQSL model's performance shows high consistency, corresponding to high stability on this dataset. These results also explain the narrow box plots obtained for the Speech Commands dataset using our proposed HQSL model, displayed in Fig. \ref{fig:feasibility}. 
\begin{table}[ht]
\centering
\captionsetup{width=\textwidth}
\caption{Raw data for Speech Commands dataset for single client-single server experiments complementing Fig. \ref{fig:feasibility}. The low standard deviations (std) relative to the means for HQSL case demonstrate high consistency in the accuracy and F1-score results}
    \begin{tabular}{ccccc}
        \hline
         & Fold & F1-Score & Accuracy & Loss \\ \hline
        \multirow{10}{*}{Classical (SL)} & 1 & 0.941733 & 0.941766 & 0.120393 \\ \cmidrule{2-5} 
        & 2 & 0.951114 & 0.951158 & 0.105335 \\ \cmidrule{2-5} 
        & 3 & 0.942385 & 0.942392 & 0.114520 \\ \cmidrule{2-5} 
        & 4 & 0.951102 & 0.951158 & 0.105450 \\ \cmidrule{2-5} 
        & 5 & 0.956093 & 0.956168 & 0.101465 \\ \cmidrule{2-5} 
        & mean & 0.948485 & 0.948528 & - \\ \cmidrule{2-5}
        & std & 0.006214 & 0.006237 & - \\ \hline
        \multirow{10}{*}{Hybrid (HQSL)} & 1 & 0.957421 & 0.957420 & 0.086337 \\ \cmidrule{2-5} 
        & 2 & 0.959869 & 0.959925 & 0.082089 \\ \cmidrule{2-5} 
        & 3& 0.961175 & 0.961177 & 0.082415 \\ \cmidrule{2-5} 
        & 4 & 0.959919 & 0.959925 & 0.086670 \\ \cmidrule{2-5} 
        & 5 & 0.959918 & 0.959925 & 0.084178 \\ \cmidrule{2-5} 
        & mean & 0.959660 & 0.959674 & - \\ \cmidrule{2-5}
        & std & \textbf{0.001368} & \textbf{0.001372} & - \\ \hline
    \end{tabular}
    \label{tab:performance_metrics_speech}
\end{table}

\bibliography{sn-bibliography, references}
\end{appendices}

\end{document}